\documentclass[12pt,letterpaper]{article}\usepackage[top=1in, left=1in, right=1in, bottom=1in]{geometry}
\usepackage[]{graphicx}\usepackage[]{xcolor}
\makeatletter
\def\maxwidth{ %
  \ifdim\Gin@nat@width>\linewidth
    \linewidth
  \else
    \Gin@nat@width
  \fi
}
\makeatother

\definecolor{fgcolor}{rgb}{0.345, 0.345, 0.345}

\usepackage{framed}
\makeatletter
 {\par\unskip\endMakeFramed%
 \at@end@of@kframe}
\makeatother

\definecolor{shadecolor}{rgb}{.97, .97, .97}
\definecolor{messagecolor}{rgb}{0, 0, 0}
\definecolor{warningcolor}{rgb}{1, 0, 1}
\definecolor{errorcolor}{rgb}{1, 0, 0}
\newenvironment{knitrout}{}{} % an empty environment to be redefined in TeX

\usepackage{alltt}
\usepackage[top=1in, left=1in, right=1in, bottom=1in]{geometry}
\usepackage[]{graphicx}
\usepackage[]{color}
\usepackage{alltt}
\usepackage{authblk} % author affiliation
\usepackage{amsmath} % align env.
\usepackage{amssymb, amsthm} % math commands
\usepackage[comma]{natbib} % bibliography: DOES NOT work with \bibliographystyle{alpha}
\usepackage{hyperref} % url & citation links
\usepackage{url} % to handle (special characters in) url's in *.bib
\usepackage{relsize} % for \mathlarger
\usepackage{enumitem} % customise bullet label & resume (don't use package enumerate}
\usepackage[final]{changes} % manual change markup
\usepackage{setspace} % for doublespacing
\allowdisplaybreaks

\usepackage{algorithm}
\usepackage{algorithmic}

\newcommand{\C}[1][]{\mathbf{C}_{#1}}
\newcommand{\I}{\mathbb{I}}
\renewcommand{\P}{\mathbb{P}}
\newcommand{\E}[1][]{\mathbf{E}_{#1}}
\newcommand{\M}[1][]{\mathbf{M}_{#1}}

\newcommand{\bsigma}[1][]{\boldsymbol{\sigma}_{#1}}
\newcommand{\x}[1][]{\boldsymbol{\xi}_{#1}}
\newcommand{\xs}[1][]{\xi^{\bsigma}_{#1}}
\newcommand{\zs}[1][]{Z^{\bsigma}_{#1}}
\newcommand{\ys}[1][]{Y^{\bsigma}_{#1}}
\newcommand{\p}[1][]{\boldsymbol{\phi}_{#1}}

\newcommand{\Y}[1][]{\mathbf{Y}_{#1}}

\newcommand{\Z}[1][]{\mathbf{Z}_{#1}}

\newcommand{\Prod}[2][]{\mathlarger\prod_{#2}^{#1}} % 2 args., 1st optional w/ empty default
\newcommand{\sumpq}{\sum_{p=1}^{n-1}\sum_{q=p+1}^{n}}

\newcommand{\prodij}{\Prod[K_n]{i=1}\Prod[K_n]{j=1}}

\newcommand{\bseta}{\boldsymbol{\eta}}
\IfFileExists{upquote.sty}{\usepackage{upquote}}{}
\begin{document}
\title{
  \textbf{A Bayesian Nonparametric Stochastic Block Model for Directed Acyclic Graphs}
}
%\author{}
\author[1]{Clement Lee}
\author[2]{Marco Battiston}
\affil[1]{School of Mathematics, Statistics and Physics, Newcastle University, UK}
\affil[2]{Department of Mathematics and Statistics, Lancaster University, UK}
%\affil[] {\texttt{clement.lee@newcastle.ac.uk}}
\maketitle

\begin{abstract}
Random graphs have been widely used in statistics, for example in network analysis and graphical models. In some applications, the data may contain an inherent hierarchical ordering among its vertices, which prevents directed edges between pairs of vertices that do not respect this order. For example, in bibliometrics, older papers cannot cite newer ones. In such situations, the resulting graph forms a Directed Acyclic Graph. In this article, we extend the Stochastic Block Model (SBM) to account for the presence of such ordering in the data, ignoring which can lead to biased estimates of the number of blocks. The proposed approach includes in the model likelihood a \emph{topological ordering}, which is treated as an unknown parameter and endowed with a prior distribution. We describe how to formalise the model and perform posterior inference for a Bayesian nonparametric version of the SBM in which both the hierarchical ordering and the number of latent blocks are learnt from the data. Finally, an illustration with real-world datasets from bibliometrics is presented. Additional supplementary materials are available online. 
\end{abstract}

\section{Introduction} \label{sect.intro}

Random graphs have been widely used in statistics and computer science over the last years to model network, interaction and bibliometric data, just to mention a few among many fields of applications. For example, in social networks, friendships or message exchanges among users can be modelled as random graphs, having users as vertices, and edges between two users if they are friends or have exchanged messages. As another example, citation datasets in bibliometrics analysis can be modelled as random graphs, having articles as vertices, and directed edges from citing articles to cited articles.

In some applications, the dataset and the corresponding graph can contain an inherent hierarchical order among its vertices. This means that directed edges can be present only from vertices that appear earlier in this order to ones that appear later. As an example, when modelling citation networks among journal articles, e.g. \citet{jj16}, articles can cite only articles that were published earlier. As another example, in a network having  the employees of a big company as vertices and directed edges from the supervisor to the supervisee, a directed path from one vertex to another can be present only if the former is higher than the latter one in the company hierarchy. Similarly, the same structure appears in a dataset having directed edges from PhD supervisors to their students. In all these applications, the observed graph is a \emph{Directed Acyclic Graph} (DAG), i.e. it cannot display any directed cycles (any directed path having the same initial and final
vertex). Moreover, it respects a latent \emph{topological ordering} among its vertices, i.e. a linear ordering $\prec$ of the vertex set $\mathcal{V}$ such that every pair $(p,q)\in \mathcal{V}\times \mathcal{V}$ can belong to the edge set $\mathcal{E}$ only if $p\prec q$. In general, the topological ordering is not unique, i.e. for a DAG there can be multiple orderings that satisfy the definition above.

When analysing a given network dataset that is a DAG, if the latent topological ordering is not properly accounted for, the statistical models can assign positive probability to `impossible' edges, i.e. directed edges in which the first vertex appears later than the second one in the topological ordering. If this is the case for many edges, the resulting estimates from the model can deteriorate dramatically. The same phenomenon arises other statistical contexts, for example in presence of structural zeros when modelling tabular data \citep{Man14}, in which assigning, whatever small, positive probability to `impossible' cells can deteriorate estimates and bias results.

In this work, we propose an approach to extending the popular \emph{Stochastic Block Model (SBM)} \citep{hll83} for random graphs in general to graphs that are specifically DAGs. The SBM, being among the most popular random graph models for vertex clustering and community detection, was initially introduced by \citet{hll83} and \citet{ww87}, then formalised as latent models by \citet{sn97} and \citet{ns01}, and has further gained popularity with the incorporation of degree correction by \citet{kn11}. In online Appendices A.1 and A.2, we provide a brief review of this model, its generalisations and approaches to posterior inference. For recent comprehensive reviews, the reader is also referred to \citet{abbe18} and \citet{lw19}, which focus on theoretical results and modelling approaches, respectively.
When extending the SBM to model a given DAG, the underlying topological ordering is unknown and not necessarily unique. Furthermore, ignoring the ordering can lead to dramatic underestimates or overestimates of the number of blocks, as the simulations in Section~\ref{sec:simulation} demonstrate. Therefore, it will be included in the likelihood of the model as a parameter, endowed with a prior and inferred a posteriori, together with the community memberships and other parameters of the model.

The background literature regarding the use of the SBM for DAGs is quite limited and mainly consists of an unpublished technical report by \cite{lw18}, which only presents a preliminary data analysis of a DAG using a SBM. Not only was degree correction not incorporated, but there was also no formal consideration of the issue of the number of blocks. Instead, the model was merely fitted with different numbers of blocks separately with no model selection, thus not justifying the interpretations of their results.

The issue of the number of communities or blocks is common with SBMs and clustering in general, and has been reviewed in Table 1 of \cite{lw19}. One common approach to this issue is the use of Bayesian nonparametrics, which assigns a prior for the allocation vector and allows the posterior sampler to infer the number of blocks from the data. \cite{ktgyu06} and \cite{sm13} initially considered the \emph{infinite relational model}, which is a variant of the plain SBM, using a Dirichlet process prior, while \cite{gbp19} extends this model by the use of a finite regime Pitman-Yor (PY) process \citep{Pit(97)}, which is reviewed in the online Appendix A.3. A further generalisation is in \cite{lrdd22}, in which the Gibbs-type prior formulation is used instead of the PY prior, for an application to criminal networks. 

In the proposed model, we also use a Bayesian nonparametric version of the SBM, but by adopting the PY process in a different way. This process allows two possible regimes: one in which the total number of blocks is finite (and unknown) and another one in which it increases with the sample size. While other works such as \cite{gbp19} have used Bayesian nonparametric formulation for the block membership, they usually focused only on the first of the two regimes. As these two regimes imply different asymptotic behaviours of the number of blocks, the model might be misspecified if only one regime is assumed before fitting to the data. To circumvent this, the choice of regime becomes another part of the proposed model, with a model selection step based on Gibbs variable selection \citep{cc95} embedded in the Markov chain Monte Carlo (MCMC) sampler.

The proposed approach is illustrated using both simulated data and real datasets from bibliometrics. In the simulation study, where the proposed model is compared to a SBM for directed graphs, which does not account for the latent order in the likelihood, the latter brings about the aforementioned issue of assigning positive probability to many `impossible' edges, as well as substantial biases in the estimates of the overall number of blocks. In the applications to two datasets of citation networks, it is shown that the model can infer the community structure and latent hierarchical order among vertices.

The rest of this article is organised as follows. The SBM for DAGs is presented in Section~\ref{sect.model}. Section~\ref{sect.application} presents both simulated (Section~\ref{sec:simulation}) and real data (Sections~\ref{sec:real.data} and \ref{sec:citation}) illustrations. A final discussion in Section~\ref{sect.discussion} concludes the article. The online appendices contain reviews of the SBM and PY process, a detailed description of the MCMC sampler to perform posterior inference for the SBM for DAG,
%including both a collapsed and an uncollapsed version of the algorithm, detailed descriptions of the split-and-merge proposal and model selection step,
additional information on the likelihood derivations, and plots from the real-data illustrations.

\section{Model} \label{sect.model}

To introduce the model and notation used throughout the rest of the paper, let us consider a directed network to be represented as a directed multi-graph $\mathcal{G}=(\mathcal{V},\mathcal{E})$, where $\mathcal{V}$ is the vertex set and $\mathcal{E}$ is the edge set. The size of $\mathcal{V}$, denoted by $n:=|\mathcal{V}|$, is the number of vertices of $\mathcal{G}$. The $n\times{}n$ adjacency matrix of the graph is denoted by $\Y:=(Y_{pq})_{1 \leq p,q\leq n}$. If there are $y$ directed edges from vertex $p$ to vertex $q$, i.e. $y$ copies of  $(p,q)$ in $\mathcal{E}$, then $Y_{pq}=y$, otherwise $Y_{pq}=0$. We also assume no self-loops, i.e. $Y_{pp}=0$ for all $p=1,2,\ldots,n$. Next, we define $\Z[n]:=(Z_1,\ldots,Z_n)$ to be the allocation vector of length $n$, where $Z_p$ is a label associated to vertex $p$. Two vertices, $p$ and $q$, belong to the same group if and only if $Z_p=Z_q$. Essentially, $\Z[n]$ represents the group memberships of the vertices. We assume there are $K_n>1$ unique labels displayed in $\Z[n]$, denoted $(Z_{1}^{*},\ldots,Z_{K_{n}}^{*})$ hence $K_n$ groups of vertices. When whether $Z_{i}^{*}$ is the label associated with vertex $p$ is of concern, for notational convenience, hereafter we write $\I(Z_{p}=i)$ in place of $\I(Z_{p}=Z_{i}^{*})$, where $\I(A)$ denotes the indicator function of the event $A$, i.e. $\I(A)=1$ if $A$ is true, $0$ otherwise. We also define $\C:=(C_{ij})_{1\leq{}i,j\leq{}K_n}$ to be the block matrix, where $C_{ij}$ represents the general connectivity between blocks $i$ and $j$. Similar to above, given $\C$ and $Z_{p}=Z_{k}^{*}$, $C_{Z_{p} j}$  will denote element $C_{kj}$.
 
The main idea of the classic SBM is that the number of edges $Y_{pq}$ from vertex $p$ to vertex $q$ is independent of that of any other dyad, \textit{conditional on their group memberships} $Z_{p}$ and $Z_{q}$ \citep{hll83,ns01}. Moreover, we will consider the \emph{degree-corrected} version of the SBM \citep{kn11}, which takes into account the degree heterogeneity of the vertices within the same group, by introducing the vector $\x=(\xi_1,\ldots,\xi_n)$ of vertex-specific parameters, where $\xi_p>0$ is the degree correction factor for vertex $p$. 
Namely, the number of edges between vertex $p$ and $q$ is modelled as
\begin{align}
Y_{pq}|\C,\Z[n],\x~\overset{\text{ind}}{\sim}~\text{Poisson}(\xi_p\xi_q C_{Z_pZ_q}).\label{eqn.directed}
\end{align}
%where $\xi_p$, $\xi_q$ and $C_{Z_pZ_q}$ are all scalars. 
%The choice of the Poisson distribution, instead of the Bernoulli distribution,  is due to a few reasons, namely, the natural extension to multi-graphs where $\Y$ can take non-negative integer values, the asymptotic equivalence between the edge probability and the expected number of edges for large sparse graphs \citep{kn11}, and also the computational simplicity that will be illustrated in our model, especially when considering the degree-corrected version of the model.

To introduce the proposed model, we restrict $\mathcal{G}$ to be a DAG for the rest of this section. To utilise a unique feature of DAGs, as discussed in Section~\ref{sect.intro}, we define $\bsigma :=(\sigma_{1},\ldots,\sigma_{n})$ as the random vector that represents the ordering of $\mathcal{G}$, with the collection of all permutations of $\{1,2,\ldots,n\}$ as the sample space. A value of $\bsigma$ is deemed topological with respect to $\mathcal{G}$ if it satisfies the definition of topological ordering, i.e. if vertex $p$ precedes vertex $q$ in the ordering, then there cannot be edges from $q$ to $p$. We define several quantities implied from $\bsigma$. First, $\p:=(\phi_1,\ldots,\phi_n)$ is the ``inverse'' of $\bsigma$, which means that if vertex $p$ is the $r$-th vertex in the topological ordering, we have $\sigma_{r}=p$ and $\phi_p=r$. Essentially, $\p$ contains the position of each vertex in $\bsigma$. Second, we define $\Z[n]^{\bsigma}:=\left(\zs[1],\zs[2],\ldots,\zs[n]\right)$ as the \textit{reordered} allocation vector, where $\zs[p]=Z_{\sigma_{p}}$ and $Z_{\sigma_{p}}$ comes from $\Z[n]$. In the same way, we define $\x^{\bsigma}:=\left(\xs[1],\xs[2],\ldots,\xs[n]\right)$ as the reordered version of $\x$. Lastly, $\Y^{\bsigma}$ is the adjacency matrix reordered by $\bsigma$ for the columns and rows of $\Y$ simultaneously, such that $Y_{pq}^{\bsigma}=Y_{\sigma_{p}\sigma_{q}}$. 

The central component of the SBM is the distribution assumption about each dyad of $\mathcal{G}$, essentially each element of $\Y$. Due to the acyclic natrue of $\mathcal{G}$, the combinations of the possible values of $(Y_{pq},Y_{qp})$ are restricted to be $(0,0)$, $(0,y)$ or $(y,0)$, for some positive integer $y\in\mathbb{N}$. Equivalently, as $\Y^{\bsigma}$ is upper triangular for a $\bsigma$ that is topological, we require that, for all dyads $(p,q)$ where $1\leq p<q\leq n$,
\begin{align}
    \ys[qp]=0,\quad\quad \ys[pq]|\C,\Z[n],\bsigma,\x& \overset{\text{ind}}{\sim} \text{Poisson}\left(\xs[p]\xs[q] C_{\zs[p]\zs[q]}\right). \label{eqn.model}
\end{align}

\subsection{Likelihood}
With the model established in Equation~\ref{eqn.model}, we can proceed to compute the likelihood. As $\Z[n]$ is unknown prior to fitting the model, it will be treated as a vector of latent variables and assigned a prior, of which the parametrisation will be detailed in Section~\ref{sect.prior_posterior}. However, given this prior choice, several quantities can be defined to facilitate the likelihood derivations. First, there will be $K_{n}$ distinct values $(Z_{1}^{*},\ldots,Z_{K_{n}}^{*})$ appearing in $\Z[n]$ with respective frequencies $\mathbf{N}:=(N_{1},\ldots,N_{K_{n}})$. The $i$-th element of $\mathbf{N}$, i.e. $N_{i}=\sum_{p=1}^n\I(Z_{p}=Z_{i}^{*})=\sum_{p=1}^n\I(Z_{p}=i)$ is then the number of vertices in group $i$. Second, we derive from $\Z[n]$ and $\Y$ (or $\Y^{\bsigma}$) two $K_{n}\times K_{n}$ matrices $\E:=(E_{ij})_{1\leq{}i,j\leq{}K_n}$ and $\M:=(M_{ij})_{1\leq{}i,j\leq{}K_n}$. The matrix $\E$ is the edge matrix between the groups, where 
\begin{align}
  E_{ij}=\sum_{p=1}^n\sum_{q=1}^nY_{pq}\I(Z_{p}=i,Z_{q}=j)=\sumpq  \ys[pq]\I(\zs[p]=i,\zs[q]=j), \label{eq:E}
\end{align} 
while in the matrix $\M$, $M_{ij}$ is the number of dyads $(p,q)$ between groups $i$ and $j$ such that vertex $p$ is topologically in front of vertex $q$. Mathematically,
\begin{align}
  M_{ij}=\sumpq\xs[p]\xs[q]\I\left(Z_{\sigma_{p}}=i,Z_{\sigma_{q}}=j\right)=\sumpq\xs[p]\xs[q]\I\left(\zs[p] =i,\zs[q]=j\right).\label{eq:M}
\end{align}

With all required quantities defined, we will derive the likelihood with observed $\Y$ given $\Z[n]$ and $\bsigma$. We first check that $\bsigma$ is topological, or equivalently $\Y^{\bsigma}$ is upper triangular, i.e. $\ys[qp]=0$ for all dyads $(p,q)$ where $1\leq p<q\leq n$, otherwise the likelihood is 0. Once $\bsigma$ is checked to be topological, using Equation~\eqref{eqn.model}, the observed data likelihood is
\begin{align} 
& \P(\Y|\C,\Z[n],\bsigma,\x) = \I(\Y^{\bsigma}\text{ upper tri.}) \times \bar{Y} \times\Prod[n]{p=1}\Prod[n]{q=1}\left(\xi_p\xi_q\right)^{Y_{pq}}\times \prodij e^{-C_{ij} M_{ij}} C_{ij}^{E_{ij}}, \label{eqn.lik_regular}
\end{align}
where $\bar{Y}=\prod_{p=1}^{n-1}\prod_{q=p+1}^{n} \left(Y^{\bsigma }_{pq}!\right)^{-1}$, and $E_{ij}$ and $M_{ij}$ are given by Equations~\ref{eq:E} and \ref{eq:M}, respectively. Detailed derivations are in the online Appendix B. Likelihood~\eqref{eqn.lik_regular} is influenced by $\Z[n]$ and $\bsigma$ through the two matrices $\E$ and $\M$.

\subsection{Priors and posterior density} \label{sect.prior_posterior}
We shall assign independent priors one by one to $\C$, $\Z[n]$, $\bsigma$ and $\x$, in order to carry out inference within the Bayesian framework. In the subsequent calculations, some additional parameters of the priors used will be included in the notation.

For $\C$, we assume each $C_{ij}$ is \textit{a priori} independent and identically distributed according to the Gamma$(a,b)$ distribution, where $a$ and $b$ are the positive shape and rate parameters, respectively. This enables $\C$ to be integrated out, to obtain
\begin{align}
    &\P(\Y|\Z[n],\bsigma,\x,a,b)
    =\mathlarger\int\P(\Y|\C,\Z[n],\bsigma,\x)\P(\C|a,b)d\C\nonumber\\
 &\quad =   \I(\Y^{\bsigma}\text{ upper tri.})\times\bar{Y}\times\Prod[n]{p=1}\Prod[n]{q=1}\left(\xi_p\xi_q\right)^{Y_{pq}}
 \times\left(\frac{b^{a}}{\Gamma(a)}\right)^{K_{n}^2}\prodij \frac{\Gamma(E_{ij}+a)}{(M_{ij}+b)^{E_{ij}+a}}. \label{eqn.lik_collapsed}
\end{align}
Independent and relatively uninformative gamma prior distributions are assigned to the parameters $a$ and $b$, as well as the components of $\x$.

For $\bsigma$, we assign a uniform prior to all permutations of $\{1,2,\ldots,n\}$, i.e. $\pi(\bsigma)=(n!)^{-1}$. There is no issue with an ordering that is not topological having a positive prior probability, as such an ordering will result in $\I(\Y^{\bsigma}\text{ upper tri.})$ and the likelihood~\eqref{eqn.lik_collapsed} being equal to 0. 

For $\Z[n]$, we assume that its components are the first $n$ elements of an exchangeable sequence $(Z_{p})_{p\in \mathbb{N}}$ driven by a \textit{Pitman-Yor (PY) process} \citep{Pit(97)}, i.e. $Z_{p}|P \stackrel{iid}{\sim}P$ for all $p\in \mathbb{N}$ and $P\sim \text{PY}(\alpha,\theta,P_{0})$. The PY process, reviewed in the online Appendix A.3, is a distribution for an unknown probability distribution $P$ and is parametrised by three hyperparameters $(\alpha, \theta, P_{0})$, where $P_{0}$, called base distribution, is a distribution on the sample space, and $\alpha$ and $\theta$ are two scalars satisfying either: 1) $0 \leq \alpha <1$ and $\theta \geq -\alpha$; 2) $\alpha <0$ and $\theta = k |\alpha |$ for $k \in \mathbb{N}$. The Dirichlet process corresponds to the special case $\alpha=0$.

\subsection{Bayesian model selection}\label{sec:model_selection}

The PY process prior for $\Z[n]$ helps select the number of blocks and distinguish between the \emph{infinite} and \emph{finite} regimes. Specifically, 
in the \emph{infinite regime},  corresponding to $0 \leq \alpha <1$ and $\theta \geq -\alpha$, $P$ has an infinite number of support points, and the sequence $(Z_{p})_{p\in \mathbb{N}}$ will display an infinite number of distinct values. In the \emph{finite regime}, corresponding to $\alpha <0$ and $\theta = k |\alpha |$ for $k \in \mathbb{N}$, $P$ has $k$ support points, which is also the total number of distinct values in $(Z_{p})_{p\in \mathbb{N}}$.  When considering the finite regime, i.e. when $\alpha <0$, we apply the re-parametrisation $(\alpha,\theta) \to (\gamma,k)$, with $\gamma:=|\alpha| >0$ and $k:=\theta / |\alpha| \in \mathbb{N}$, and assign a prior to $(\gamma,k)\in \mathbb{R}_{+}\times \mathbb{N}$. 

As detailed in the online Appendix A.3, it is possible to compute the marginal likelihood (when $P$ is marginalised out) of a sample $\Z[n]$ driven by the $\text{PY}$ process, which we denoted by $\P(\Z[n]|\bseta_r)$, where $\bseta_r$ is a parameter vector of length 2, dependent on the choice of regime $r\in\{0,1\}$, either infinite $\bseta_0=(\alpha,\theta)$ or finite $\bseta_1=(\gamma,k)$. This in turn requires the specification of the prior of $\bseta_r$ under both regimes. Under the infinite regime, $r=0$ and $\bseta_r=\bseta_0=(\alpha,\theta)$, and we assume that $\alpha\sim\text{Uniform}[0,1]$ and $\theta+\alpha$ follows a Gamma distribution. This choice of dependent priors for $\alpha$ and $\theta$ is done in order to include all possible values in parameter space. %Indeed, when $0<\alpha<1$, the range of possible values of $\theta$ is $\theta>-\alpha$. 
Under the finite regime, $r=1$ and $\bseta_r=\bseta_1=(\gamma,k)$, and we assume that $\gamma$ and $k$ are independent \textit{a priori}, $\gamma$ follows a Gamma distribution, and $k$ follows a truncated negative binomial distribution, with parameters $a_k$ and $b_k$, and density
\begin{align}
    \P(k =k'|a_k,b_k)=\left(1-b_k^{a_k}\right)^{-1}\times\frac{\Gamma(k'+a_k)}{\Gamma(a_k) k'!} ~b_k^{a_k} (1-b_k)^{k'},\qquad k'=1,2,\ldots \nonumber
\end{align}
where $\Gamma(\cdot)$ is the gamma function. The factor $(1-b_k^{a_k})^{-1}$ is due to the truncation of $0$ from the original support of the negative binomial distribution. The negative binomial is over-dispersed, allowing possibly high variance, hence high prior uncertainty of the value of~$k$. 

The regime-dependent parameters and their priors are introduced this way because, ultimately, we want to enable model selection of the regime, which in turn requires the prior of $r$, denoted by $\P(r)$. The boundary cases $\P(r=0)=1$ and $\P(r=1)=1$ represent staying within the infinite and finite regimes, respectively, while model selection takes place when $0<\P(r=0)<1$, allowing the data to select the more suitable among the two regimes. The model selection step, implemented within the MCMC sampler, follows the algorithm of \cite{cc95}. Moreover, the prior choice for $r$  allows unbalanced weights between models, as this may improve mixing of the model selection step, as shown in \cite{fp08}, in the context of reversible jump MCMC. At the end of Section~\ref{sec:simulation}, in Table~\ref{tab:BayesFactor}, we show results from a simulated study in which the posterior sampler in general recovers the correct regime between the finite and infinite ones. A detailed description of the model selection step can be found in the online Appendix C.2.

As all priors required have been specified, the joint posterior of $\Z[n]$, $\bsigma$, $\bseta_r$, $a$ and $b$ (and $r$), up to a proportionality constant, is
\begin{align}
&\P(\Z[n],\bsigma,\x,\bseta_r,a,b,r|\Y)\propto
\P(\Y|\Z[n],\bsigma,\x,a,b)\P(\Z[n]|\bseta_r)\P(\bsigma)\P(\x)\P(\bseta_r|r)\P(a)\P(b)\P(r). \label{eq:posterior}
\end{align}
We sample from this joint posterior using MCMC, of which the detailed description is available in the online Appendix C.1.

\subsection{Posterior point estimate of $\Z[n]$}
To provide a posterior point estimate of $\Z[n]$, we follow the clustering approach introduced in \citet{Mei(07)} and further discussed in \citet{Wad(18)}. Specifically, the point estimate, denoted by $\hat{\Z}_n$, is obtained using a decision theoretic approach, by minimizing with respect to the posterior distribution a loss function on the space of allocation vectors,
\begin{align} \label{eq:point.estimate}
\hat{\Z}_n= \underset{{\tilde{\Z}_n}}{\text{argmin}}~\mathbb{E} \left[ L(\Z[n],\tilde{\Z}_n) | \mathbf{Y} \right] = \underset{{\tilde{\Z}_n}}{\text{argmin}}\sum_{\Z[n]}  L(\Z[n],\tilde{\Z}_n) \P(\Z[n]|\Y).
\end{align}
For the loss function $L(\Z[n],\tilde{\Z}_n)$, \citet{Mei(07)} chose the \emph{Variation of Information} (VI), defined as
\begin{align*}
    \text{VI}(\Z[n],\tilde{\Z}_n)  
     = \sum_{i=1}^{K_{n}}\frac{n_{i+}}{n} \log \left(\frac{n_{i+}}{n} \right) + \sum_{j=1}^{\tilde{K}_{n}} \frac{n_{+j}}{n} \log \left(\frac{n_{+j}}{n} \right) - 2 \sum_{i=1}^{K_{n}} \sum_{j=1}^{\tilde{K}_{n}}\frac{n_{ij}}{n} \log \left(\frac{n_{ij}}{n} \right),
\end{align*}
where $n_{ij}=\sum_{p=1}^{n} \mathbb{I}(Z_{p}=i,\tilde{Z}_{p}=j)$, $n_{i+}=\sum_{j=1}^{\tilde{K}_{n}}n_{ij}$, and $n_{+j}=\sum_{j=1}^{K_{n}}n_{ij}$.
The loss function $L(\Z[n],\tilde{\Z}_n)$ can be seen as a distance between $\Z[n]$ and $\tilde{\Z}_n$, which can be computed even if $K_n\neq \tilde{K}_n$, i.e. the numbers of groups implied by $\Z[n]$ and $\tilde{\Z}_n$ are different.

\subsection{Identifiability of $\Z[n]$ and $\bsigma$}

%While the likelihood is invariant to permutations of the labels in $\Z[n]$, the identifiability issue is resolved through obtaining the posterior point estimate of $\Z[n]$. This will be implemented in our application to one of the data sets, towards the end of Section~\ref{sec:real.data}. On the other hand, the likelihood is not invariant to permutations of $\bsigma$ as it is an ordering. Therefore, there are no identifiability issues for $\bsigma$, as are shown in the results in Section~\ref{sec:real.data}.

Our inference approach does not raise any concern of identifiability issues. For $\Z[n]$, the problem of label switching in Bayesian MCMC estimation of finite mixture models and related models is indeed less common in nonparametric models, given that the prior of the block weights is not symmetric \citep{jhs05,pr08}. This issue is further avoided through obtaining the posterior point estimate of $\Z[n]$ defined in Equation~\ref{eq:point.estimate}, which minimizes the posterior expectation of a well-chosen loss function. For $\bsigma$, or equivalently $\p$, the likelihood is not invariant to its permutations as it is an ordering. The absence of identifiability issues for $\bsigma$ is further evidenced in our applications in Section~\ref{sect.application}, where the posterior estimates closely approximate the known true topological ordering in the simulated examples (Figure~\ref{fig:sim_topo}), and the posterior distributions of $\p$ do not look particularly flat for a real dataset (Figure~\ref{fig:sna_topo}).

%Results from the experiments of Section~\ref{sect.application} do not raise any concern of identifiability issues of $\Z[n]$ and $\bsigma$. With regard to $\Z[n]$, trace plots do not show any evidence of label switching. This latter problem is indeed less common in non-parametric models, given that the prior of the block weights is not symmetric. With regard to $\bsigma$, in both simulated and real data examples, there were no signs of unidentifiability of the ordering of the vertices. In the simulated examples of Section~\ref{sec:simulation}, where the true topological ordering is known, the posterior estimates obtained by the algorithm result in very close approximations to the true ordering. On the citation dataset of Subsection~\ref{sec:citation}, running the algorithm with different seeds or different initialisations produces robust estimates of the ordering, with the same articles %(those which are highly cited, or papers cited by highly cited ones) 
%on top of the ordering. 
%Moreover, Figures~\ref{fig:sim_topo} and~\ref{fig:sna_topo}, which summarise the posterior distributions of the topological ordering in simulated and real data respectively, do not suggest that the posterior distribution is particularly flat, with many parameter values having similar posterior mass. Finally, in Appendix D of the online supplementary material, trace-plots of the topological ordering are shown to mix well.

\section{Application} \label{sect.application}

\subsection{Simulated data} \label{sec:simulation}

This section presents an illustrative simulation study in which the proposed DAG-SBM (Equation~\eqref{eqn.model}) is compared with the SBM for a directed graph (Equation~\eqref{eqn.directed}, referred to as the directed SBM). The purpose of this study is to show how fitting the directed SBM without considering the inherent hierarchical ordering among vertices can result in assigning positive posterior probability to many ``impossible" edges, and substantial bias in the estimate of the overall number of blocks in the model.

Upon sampling the parameters from their priors specified in Section~\ref{sect.prior_posterior}, adjacency matrices are sampled from the DAG-SBM for different sample sizes $n=\{250$,$500$,$1000\}$ and combinations of the PY parameters, $(\alpha,\theta)=\{(-1,10)$, $(0,1)$, $(0.1,1)$, $(0.6,1)\}$, thus including both finite and infinite regimes. Both the DAG-SBM and directed SBM are fitted to the resulting adjacency matrices, also under both regimes. 

\begin{knitrout}
\definecolor{shadecolor}{rgb}{0.969, 0.969, 0.969}\color{fgcolor}\begin{figure}[!htbp]

{\centering \includegraphics[width=0.29\linewidth]{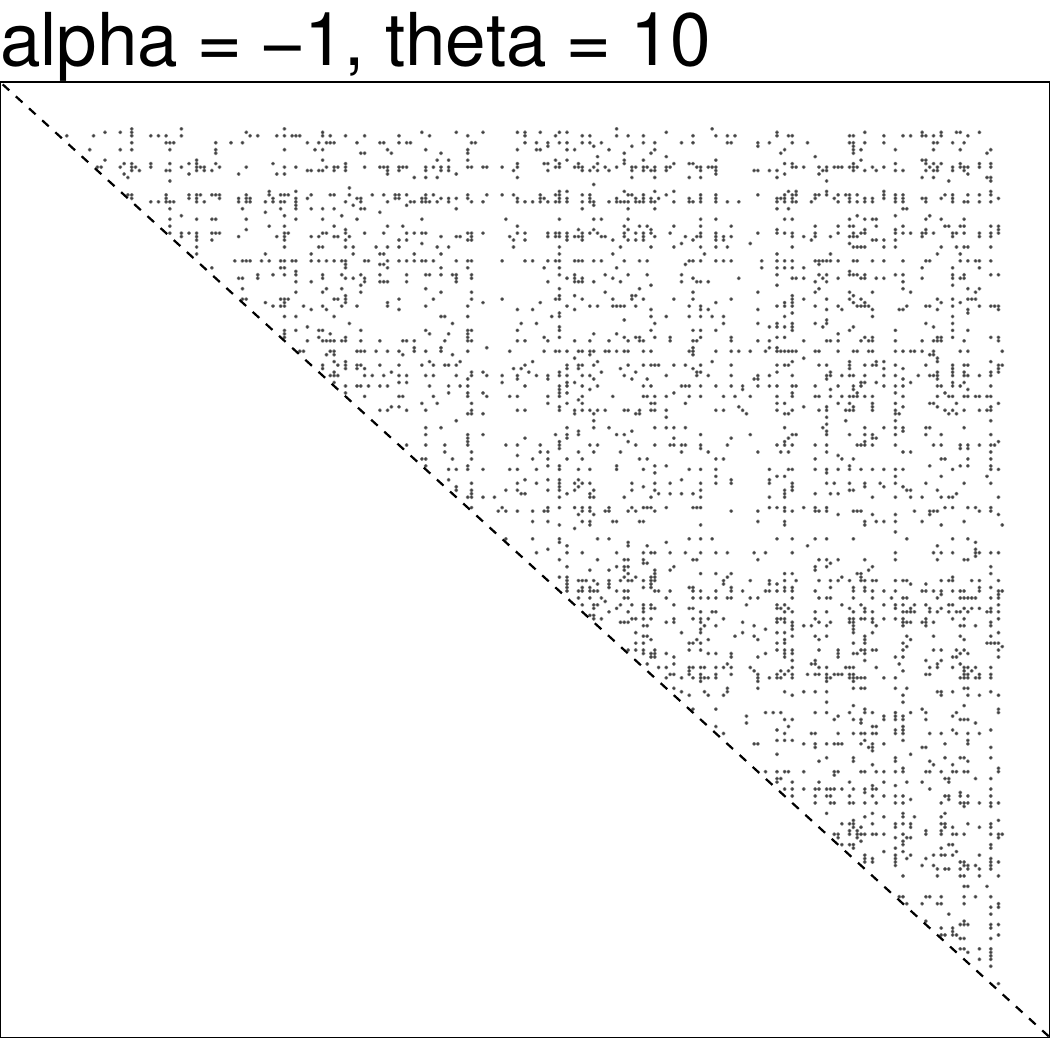} 
\includegraphics[width=0.29\linewidth]{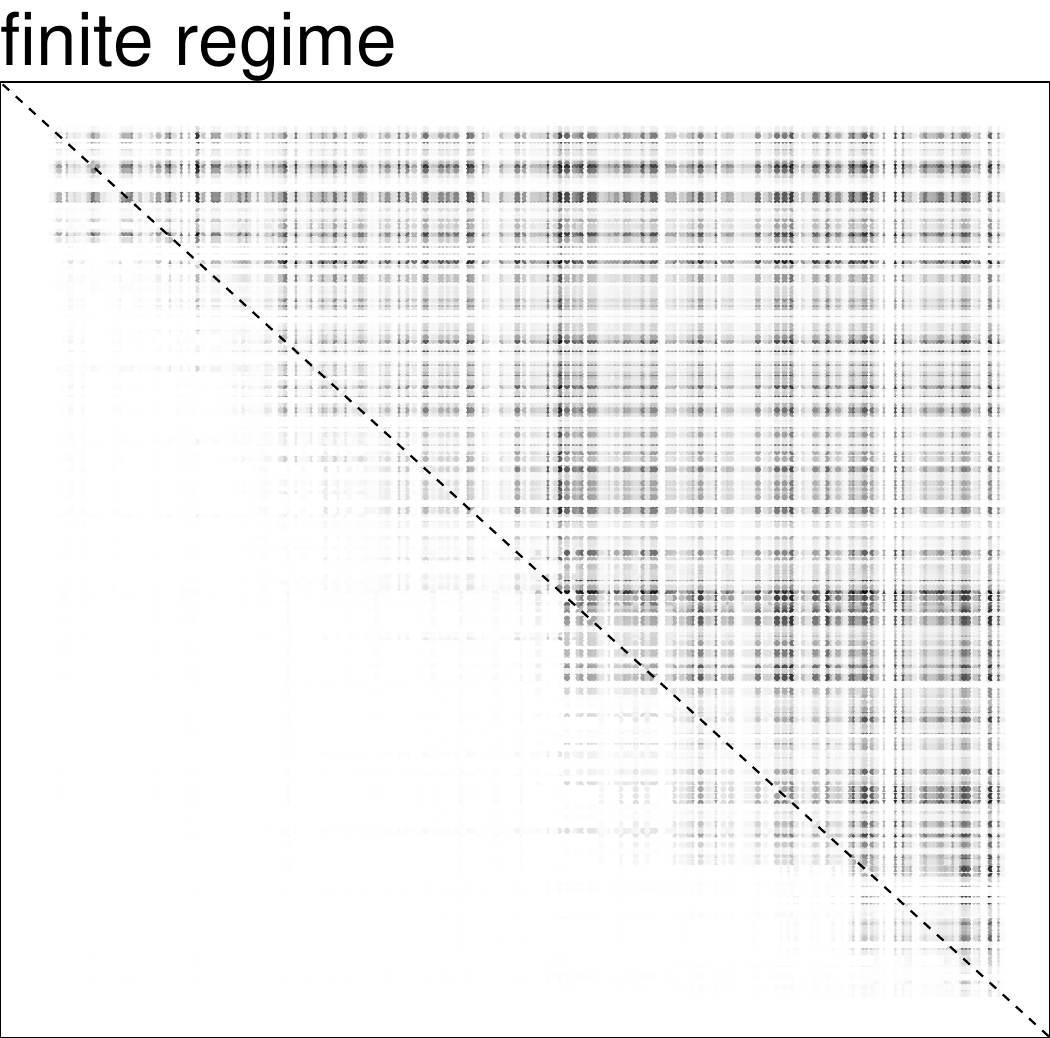} 
\includegraphics[width=0.29\linewidth]{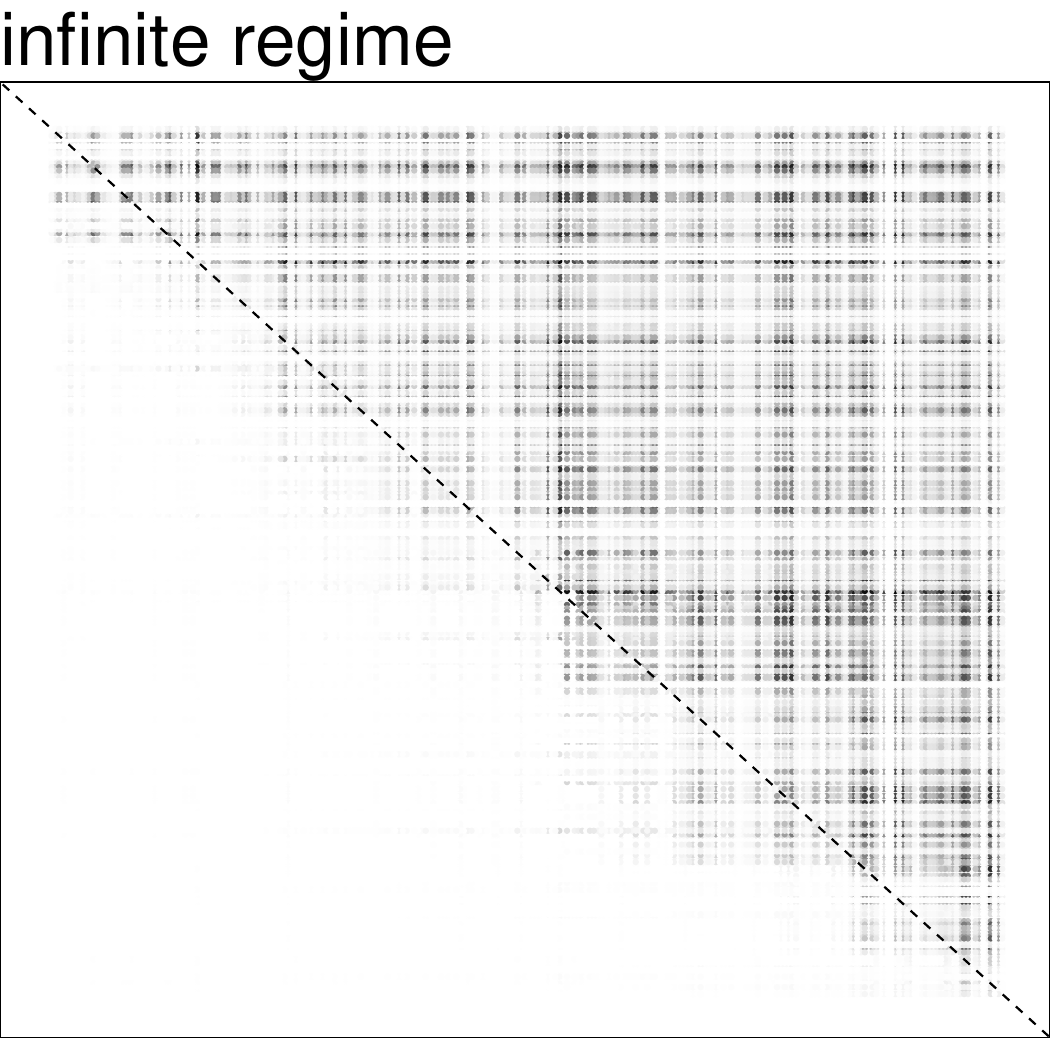} 
\includegraphics[width=0.29\linewidth]{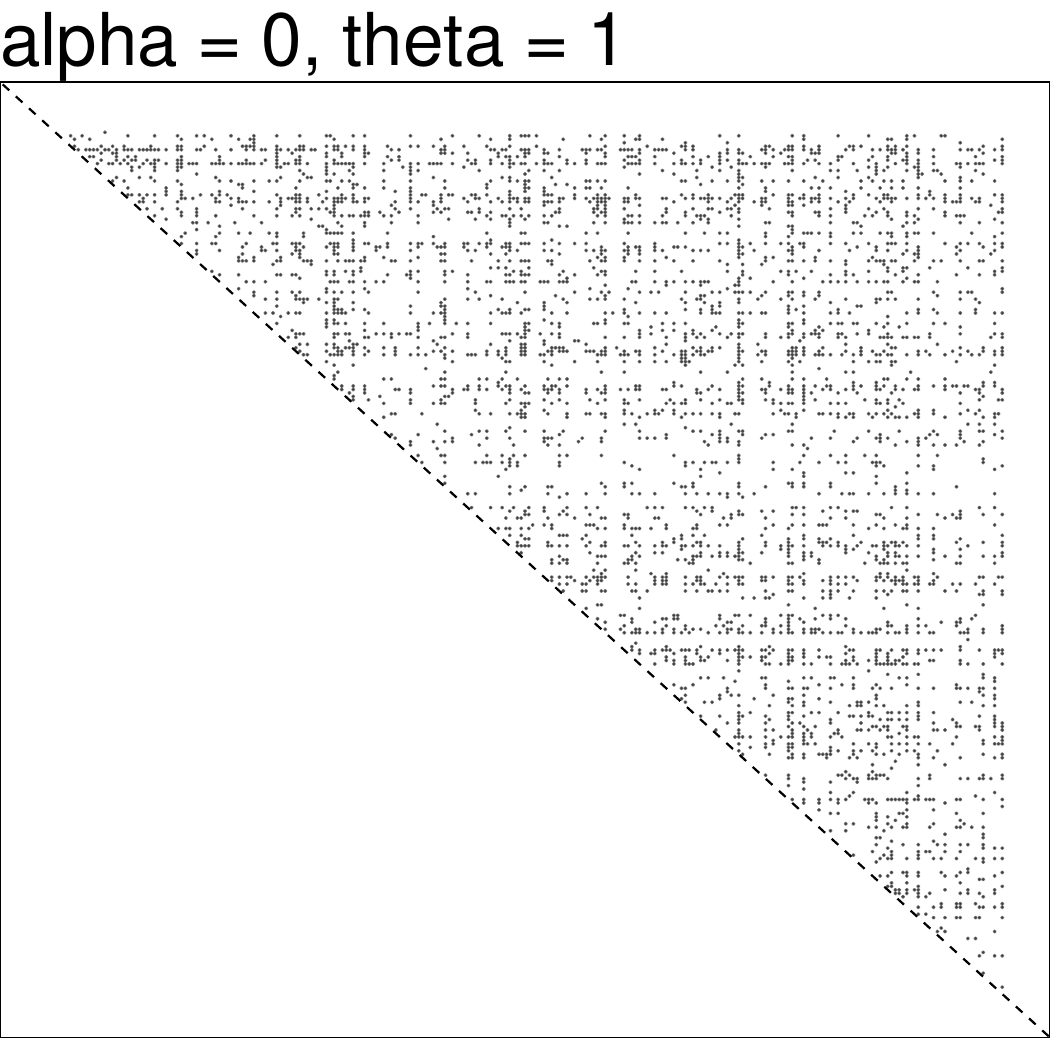} 
\includegraphics[width=0.29\linewidth]{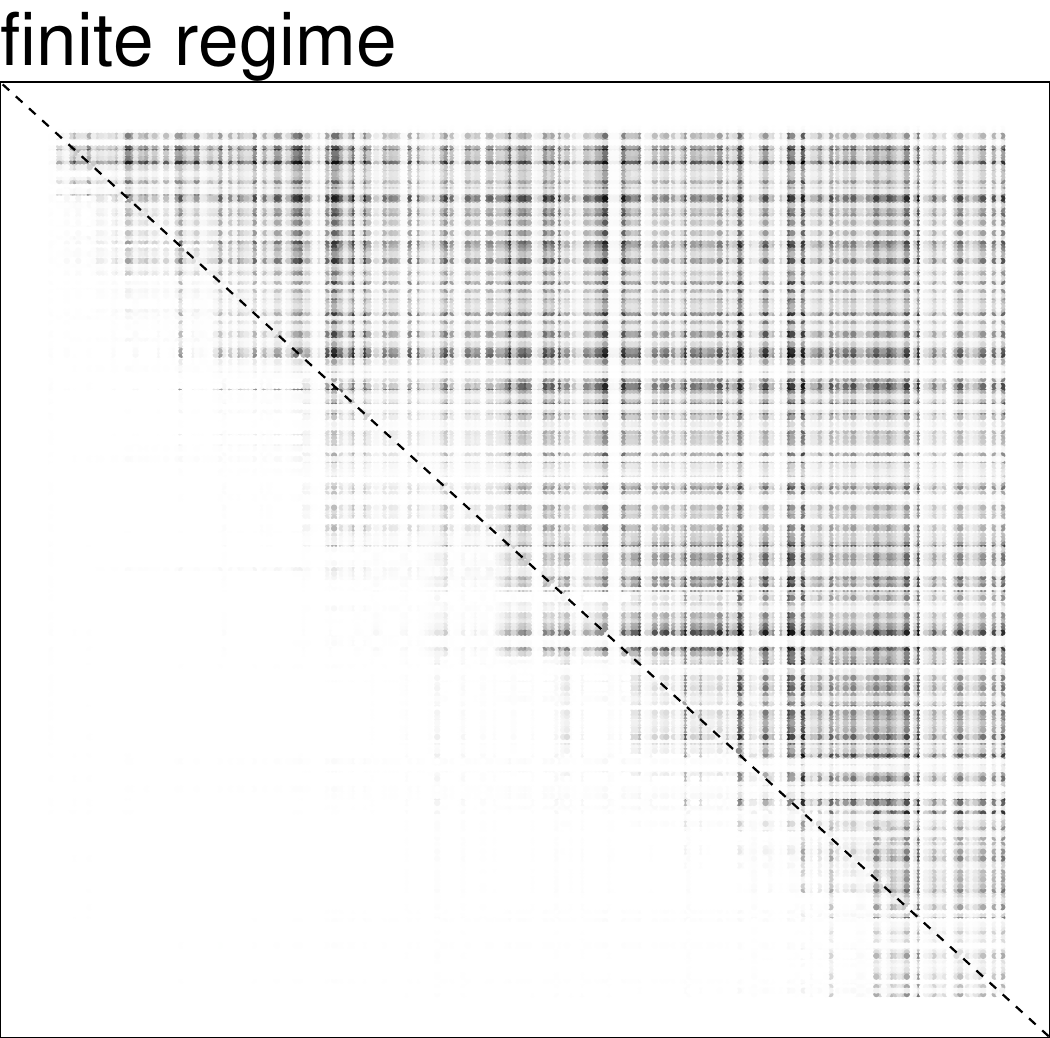} 
\includegraphics[width=0.29\linewidth]{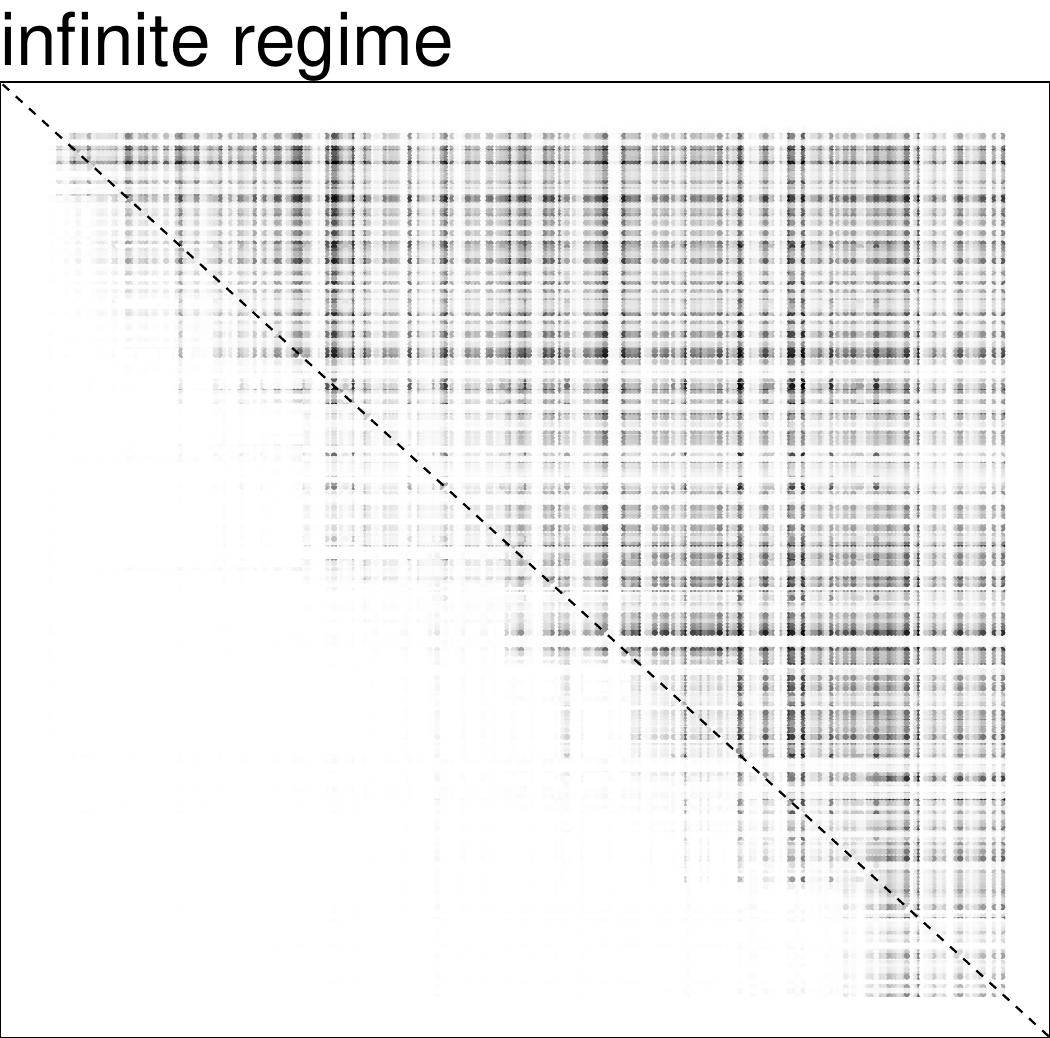} 
\includegraphics[width=0.29\linewidth]{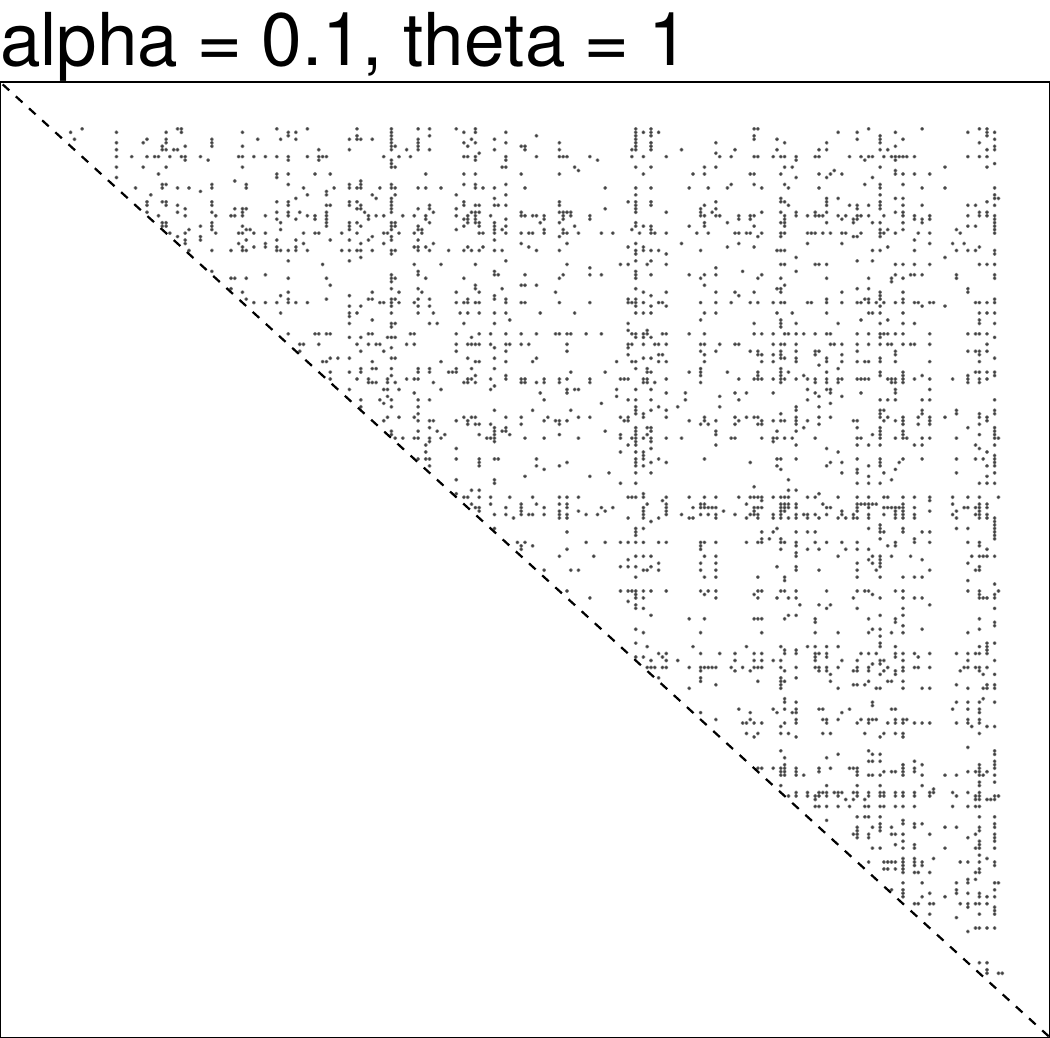} 
\includegraphics[width=0.29\linewidth]{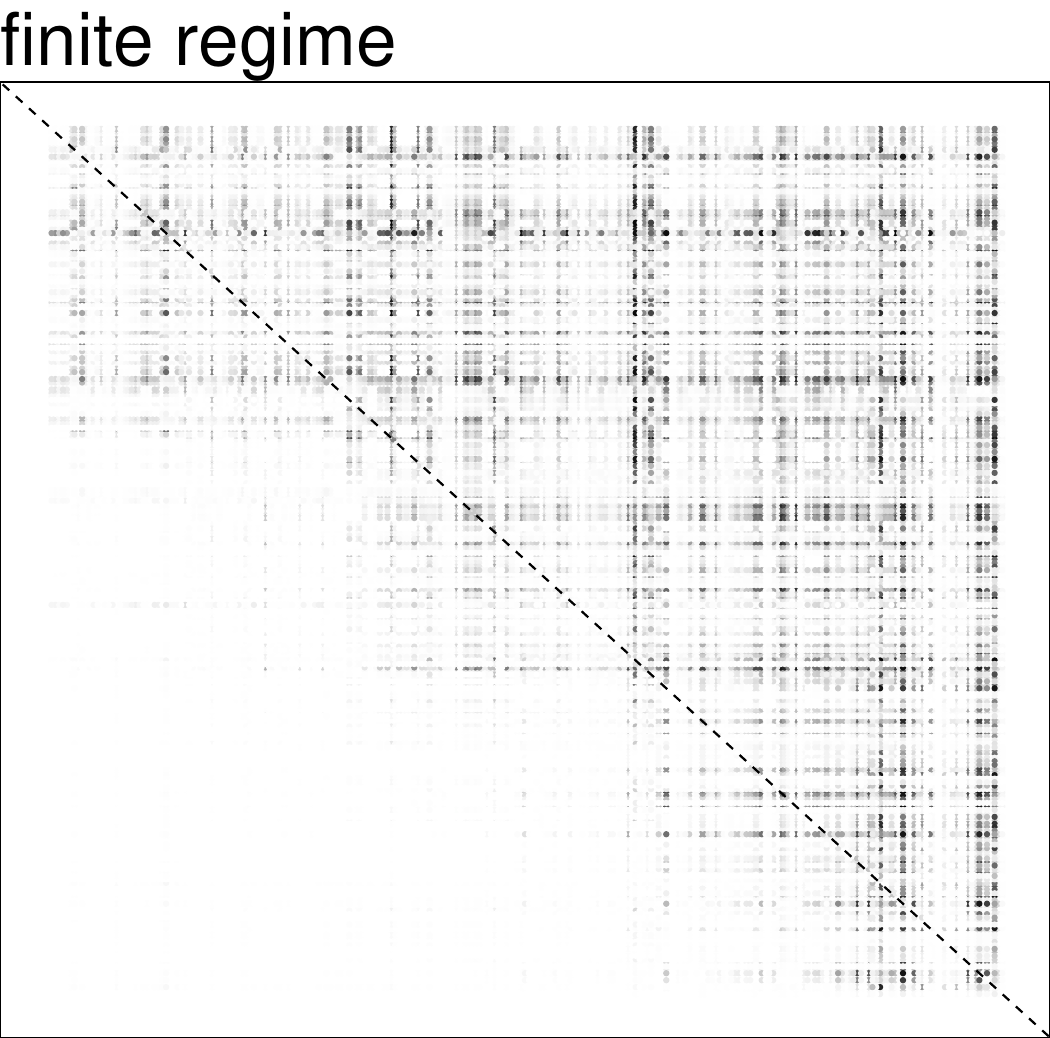} 
\includegraphics[width=0.29\linewidth]{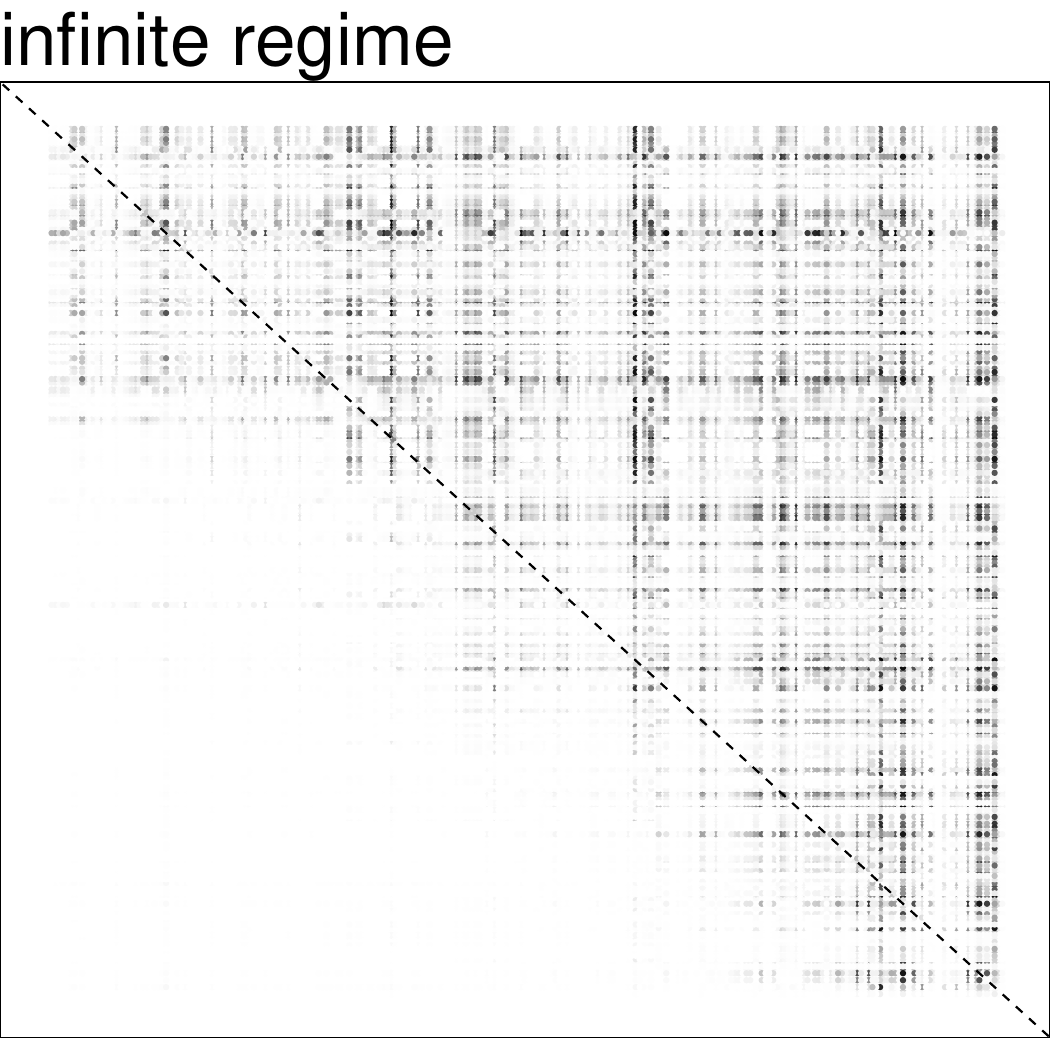} 
\includegraphics[width=0.29\linewidth]{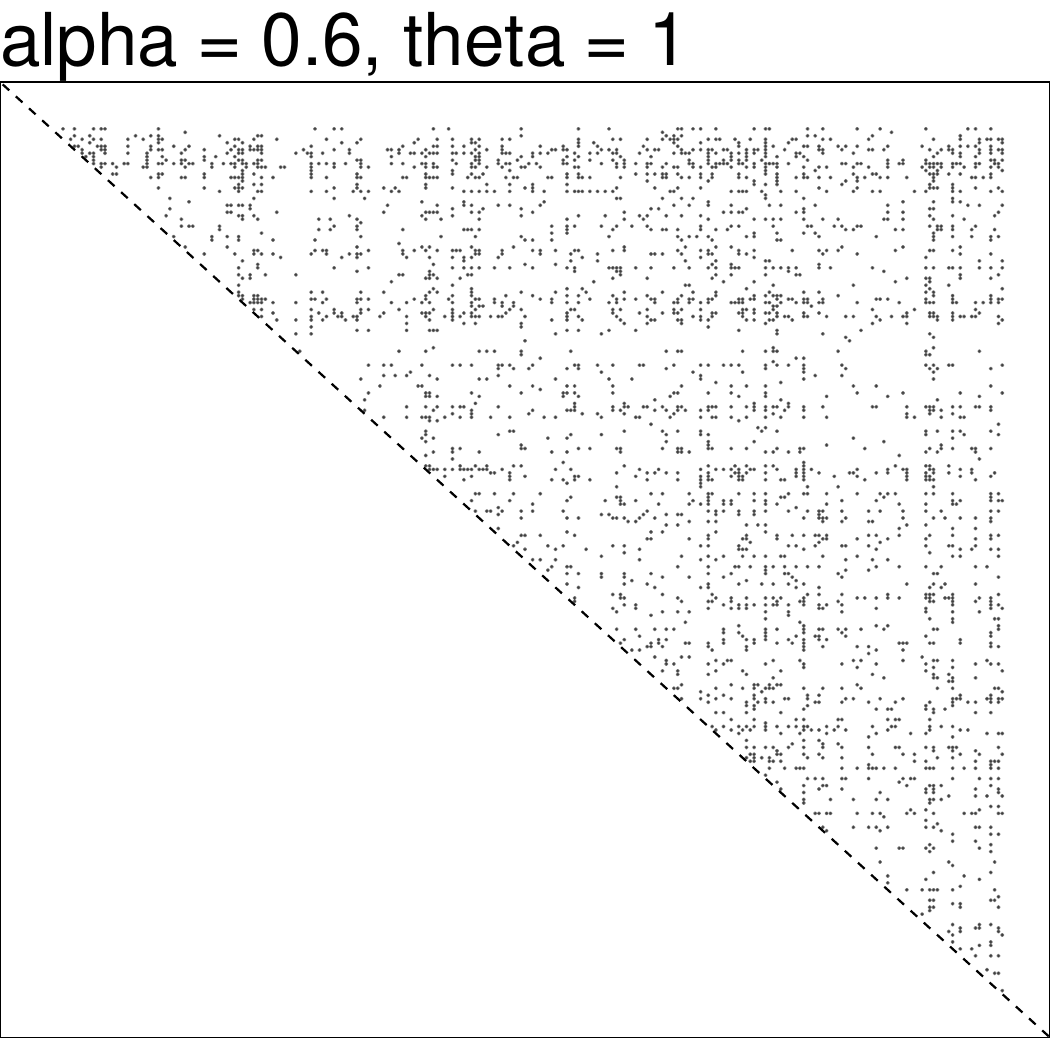} 
\includegraphics[width=0.29\linewidth]{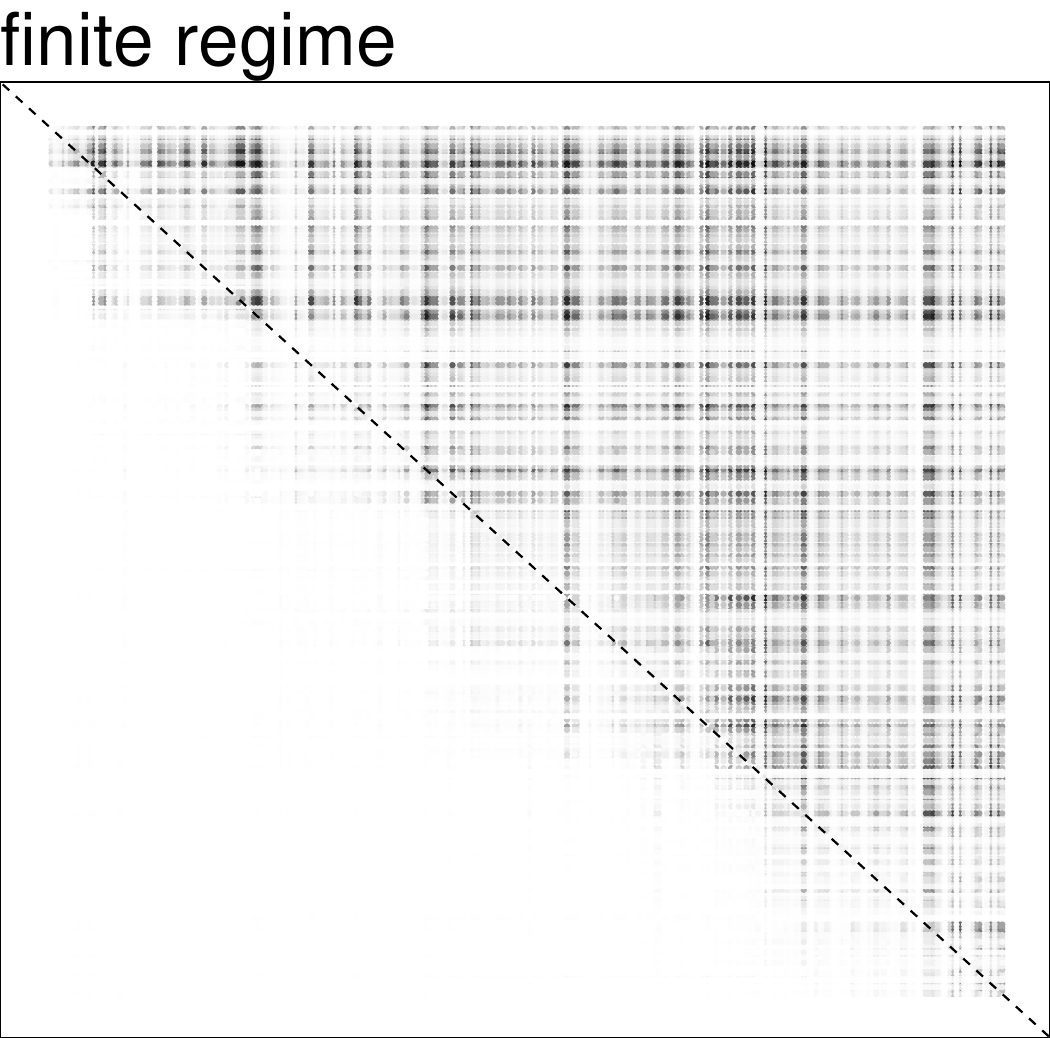} 
\includegraphics[width=0.29\linewidth]{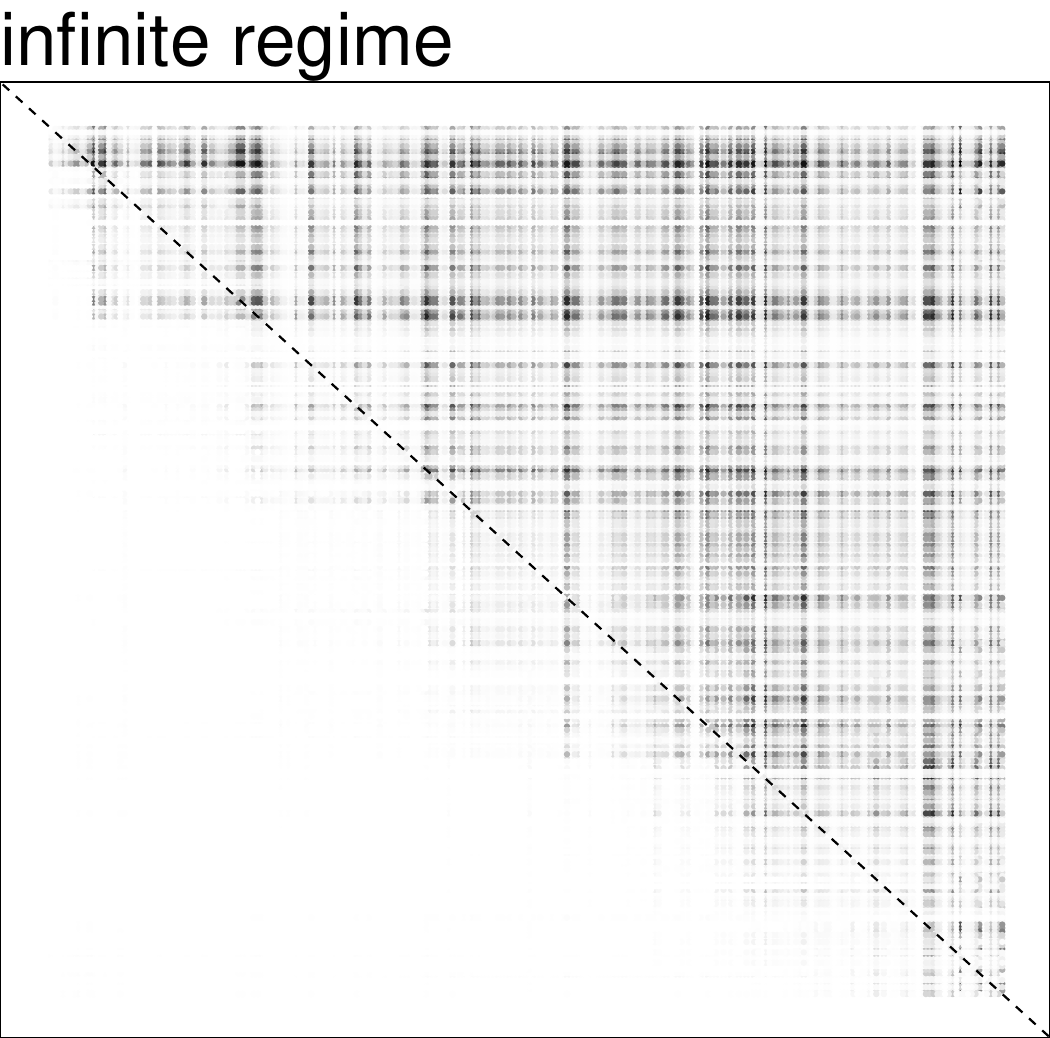} 

}

\caption{Actual adjacency matrix (left) for data simulated from the DAG-SBM (Equation~\eqref{eqn.model}) with $n=250$ and different values of $(\alpha,\theta)$, and estimated probabilities by the directed SBM (Equation~\eqref{eqn.directed}) under the finite (middle column) and infinite (right) regime, ordered by true $\bsigma$. The middle and right plots correspond to the results in finite and infinite regimes in the corresponding bottom left panels of Figure~\ref{fig:sim_plot_K}. The dashed line is the major diagonal.}\label{fig:sim_plot_A}
\end{figure}

\end{knitrout}

Figure~\ref{fig:sim_plot_A} plots the actual adjacency matrix $\Y^{\bsigma}$ with $n=250$ (not shown for $n=500$ and $n=1000$), ordered according to the true topological ordering, alongside with the posterior point estimates of $\Y^{\bsigma}$, for each simulated dataset fit by the directed SBM. The true adjacency matrices in the first column are necessarily upper triangular, as they are reordered according to the true $\bsigma$. However, from the second and third columns, it is evident that many edges in the lower triangular sections have positive posterior probability, while it should be equal to zero. This is especially true for vertices having either high in-degree or out-degree counts. This is due to the fact that the directed SBM is row and column exchangeable, and vertices with high in-degrees tend to also have high out-degree counts. However, when a hierarchical ordering is present, the model is not exchangeable anymore and the out- and in-degrees of each vertex are negatively correlated, with vertices at the beginning of $\bsigma$ having higher out-degree than in-degree counts, and the opposite for vertices at the end of $\bsigma$.

\begin{knitrout}
\definecolor{shadecolor}{rgb}{0.969, 0.969, 0.969}\color{fgcolor}\begin{figure}

{\centering \includegraphics[width=0.9\linewidth]{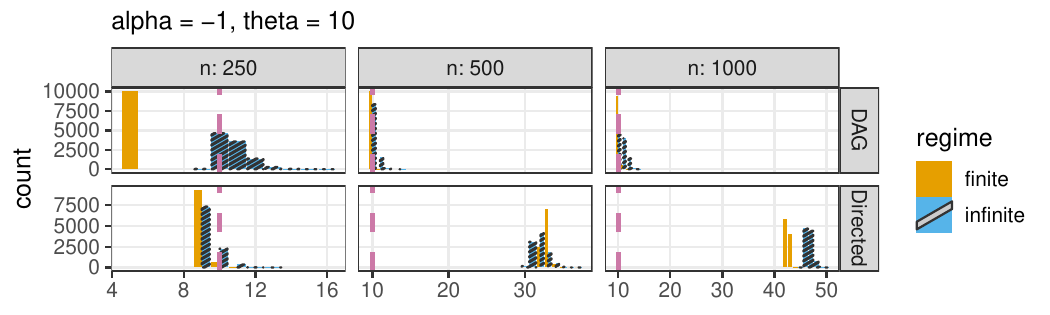} 
\includegraphics[width=0.9\linewidth]{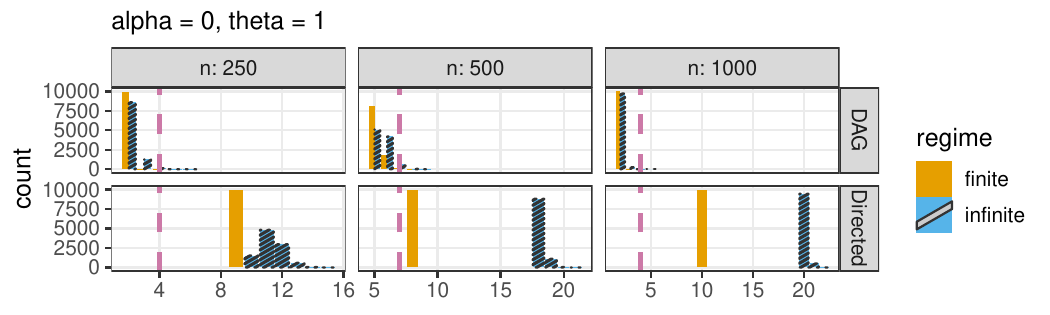} 
\includegraphics[width=0.9\linewidth]{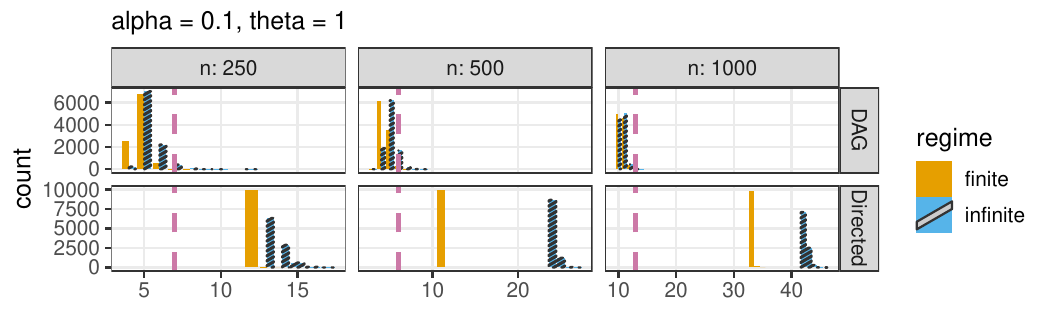} 
\includegraphics[width=0.9\linewidth]{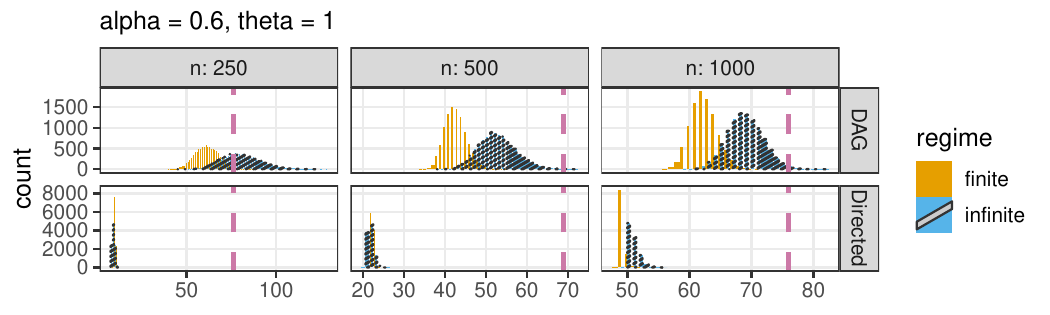} 

}

\caption{Posterior histogram of $K_n$ for data simulated from the DAG-SBM (Equation~\eqref{eqn.model}), with different values of $(\alpha,\theta,n)$, fitted by both the DAG-SBM (top row) and directed SBM (Equation~\eqref{eqn.directed}, bottom row), under the finite (solid) and infinite (stripe) regimes. The vertical dashed line is the true value of $K_n$.}\label{fig:sim_plot_K}
\end{figure}

\end{knitrout}

Figure~\ref{fig:sim_plot_K} displays the posterior histograms of the number of blocks $K_{n}$ estimated using the DAG-SBM and the directed SBM, for the same sets of simulated data. On one hand, the directed SBM, which does not include the topological ordering in the likelihood, has the tendency of heavily underestimating or overestimating $K_n$. The increase in the number of blocks also negatively affects the computational costs of the algorithm, compared to the DAG-SBM. A comparison of the computational times between the two models is provided in the online Appendix D.1. On the other hand, the DAG-SBM provides reasonable estimates of $K_{n}$ across different sample sizes and for both regimes of the PY parameters. Furthermore, for the DAG-SBM, the model selection step in the MCMC sampler successfully recovers the true regime of the PY parameters. Indeed, Table~\ref{tab:BayesFactor} displays the estimated Bayes factor for the model selection step between the finite and infinite regime, for a range of different values of the hyperparameters of the PY process, including both logarithm (corresponding to the Dirichlet Process, $\alpha=0$) and polynomial growth ($0<\alpha<1$) for $K_{n}$.
The estimated Bayes factor suggests that the posterior sampler is capable of recovering the correct regime from the data, particularly when $\alpha$ is far from $0$, and being generally indifferent between the regimes when $\alpha$ is close to $0$, and $K_{n}$ grows logarithmically, hence very slowly.

\begin{table}[t!]
    \centering
    \begin{tabular}{cc|c|ccc}
    \hline
    $\alpha$     & $\theta$     & \text{Regime} & $n=250$ & $n=500$    & $n=1000$ \\
    \hline
     $-1$    & $10$  & \text{Finite} & $8.30$   &  $612.58$ &  $2.72$       \\
     $0$ &  $1$ & \text{Infinite: Logarithm growth} & $2.56$ &  $0.85$ &  $19.12$           \\
    $0.1$ & $1$ & \text{Infinite: Polynomial growth} & $1.11$ &  $0.40$ &  $0.89$           \\
    $0.6$ & $1$ & \text{Infinite: Polynomial growth} & $7.93\times10^{-4}$ &  $7.62\times10^{-7}$ & $1.92\times10^{-11}$           \\
    \hline
    \end{tabular}
    \caption{Estimated Bayes factor for the finite regime to the infinite regime when applying the model selection in the MCMC sampler for the DAG-SBM to the simulated data, for different values of $(\alpha,\theta,n)$. A value above (below) $1$ favours the finite (infinite) regime.}
    \label{tab:BayesFactor}
\end{table}

Finally, we look at the posterior of $\bsigma$, or equivalently the positions of the vertices in the topological ordering, $\p$. The mixing of the MCMC is good, and the trace plots are available for some components of $\p$ in the online Appendix D. The posterior density of each component of $\p$ is plotted as a row in Figure~\ref{fig:sim_topo}, with the rows in the true topological ordering. The concentration of posterior mass on the main diagonal suggests that this latent ordering can be well recovered without identifiability issues. %The absence of identifiability issues is also evidenced in the trace plots in the online Appendix D, where the posterior for different components of $\p$ spans different ranges (but is highly similar between the two regimes).

\begin{knitrout}
\definecolor{shadecolor}{rgb}{0.969, 0.969, 0.969}\color{fgcolor}\begin{figure}

{\centering \includegraphics[width=0.48\linewidth]{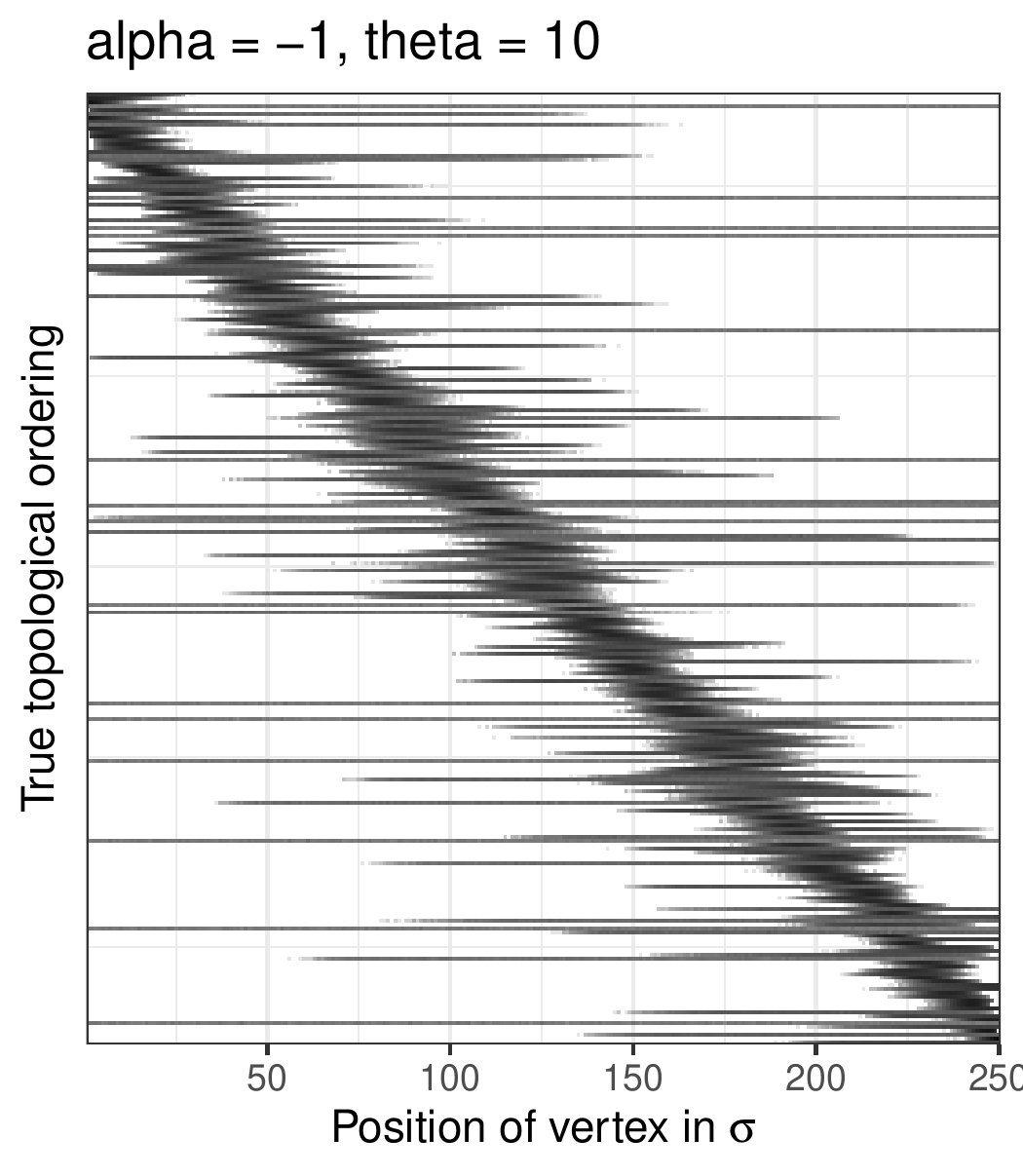} 
\includegraphics[width=0.48\linewidth]{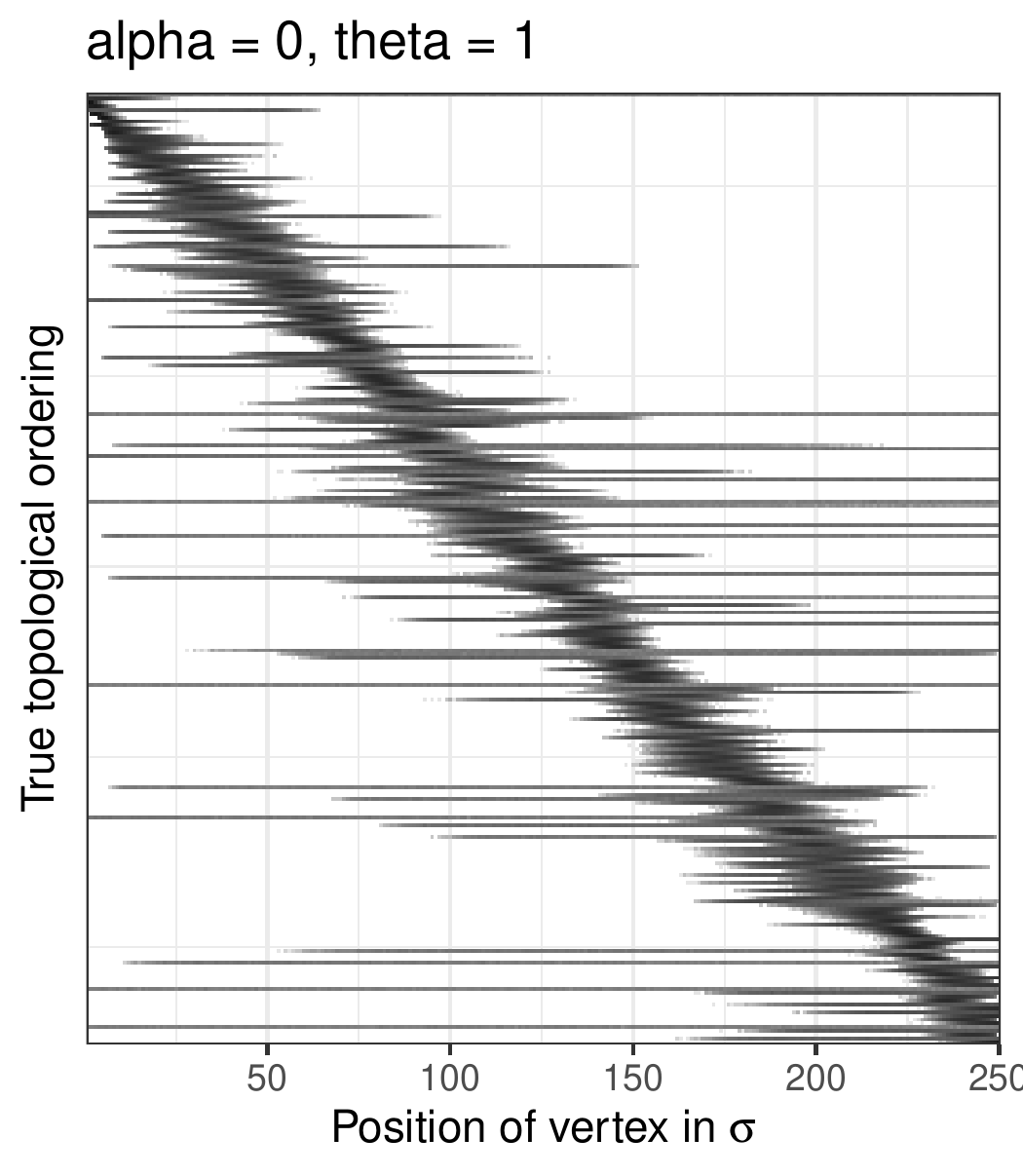} 
\includegraphics[width=0.48\linewidth]{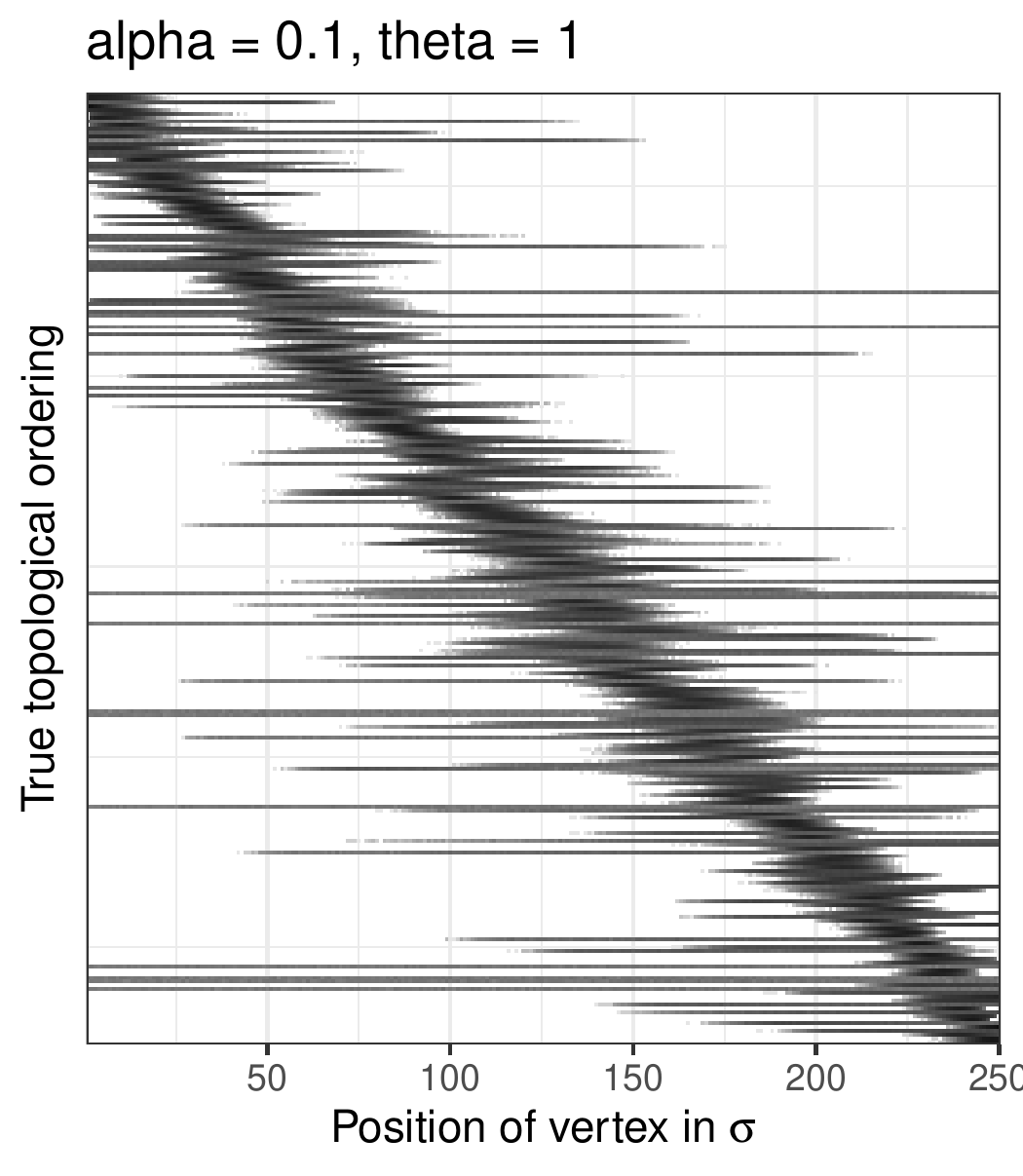} 
\includegraphics[width=0.48\linewidth]{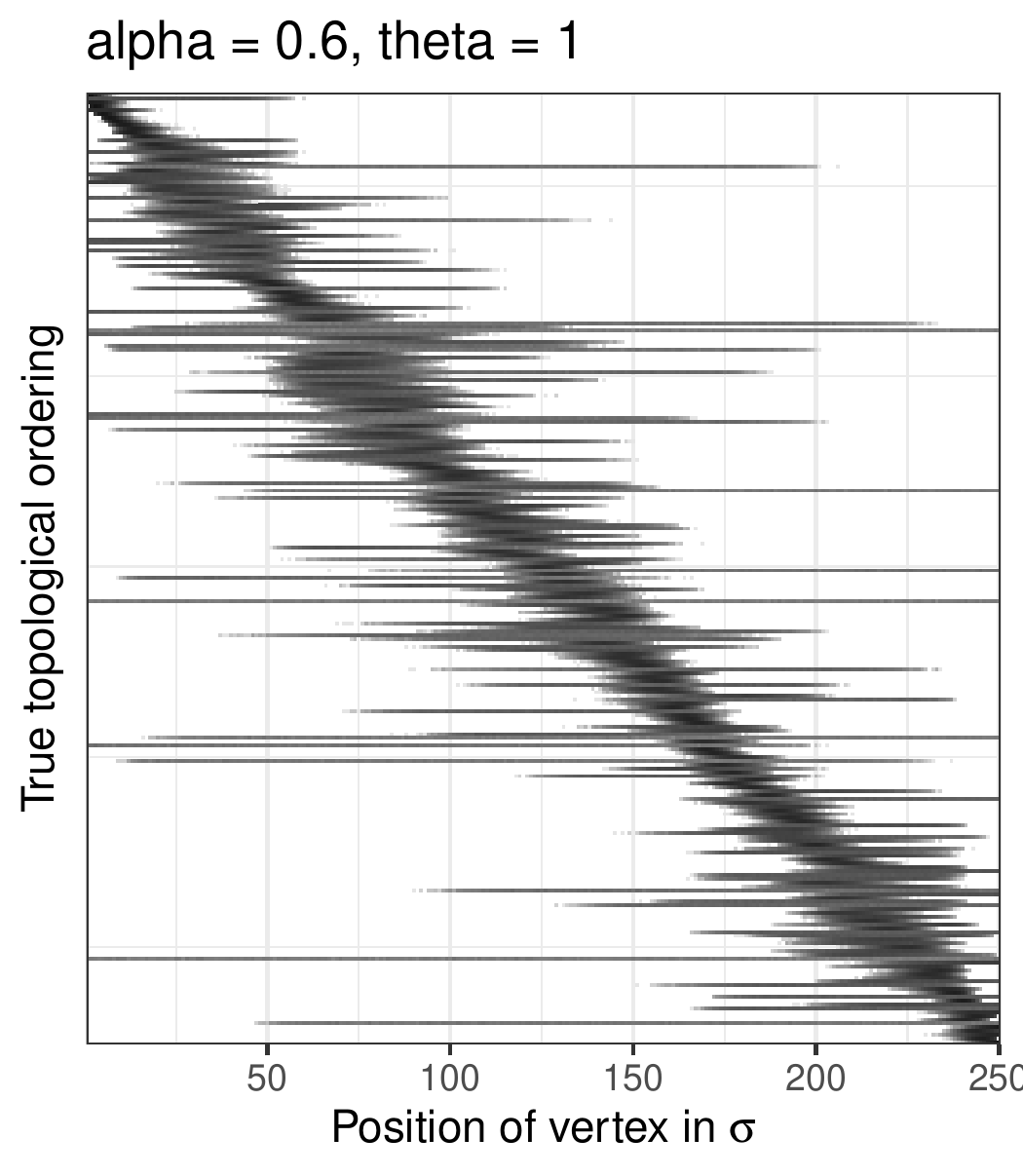} 

}

\caption{The posterior density of the positions of the vertices in $\bsigma$ for $n=250$ and different values of $(\alpha,\theta)$ for data simulated from the DAG-SBM (Equation~\eqref{eqn.model}), fitted by the DAG-SBM under the true regime according to $\alpha$. Each row represents the posterior density of a component of $\p$.}\label{fig:sim_topo}
\end{figure}

\end{knitrout}

\subsection{SNA citation network} \label{sec:real.data}
Next, we look at the citation network analysed by \cite{lw18}. It contains 1118 citations (edges) between 135 articles (vertices) which are related to social network analysis (SNA). We shall call this data the SNA citation network hereafter.

The MCMC sampler for the DAG-SBM outlined in the online Appendix C was applied, with 20000 iterations obtained after a burn-in period of 100000 and thinning of 2000, i.e. keeping 1 sample every 2000 iterations, with $a\sim\text{Gamma}(1,0.01)$, $b\sim\text{Gamma}(1,0.01)$. Such settings were used three times, once assuming the infinite regime $(r=0)$ with $\theta+\alpha\sim\text{Gamma}(1,0.01)$, once assuming the finite regime $(r=1)$ with $\gamma\sim\text{Gamma}(1,0.01)$ and $k\sim$ truncated negative binomial$(1,0.01)$, and once for model selection with $\P(r=1)=0.2$, i.e. $\P(r=0)=0.8$. All three runs were performed on a Linux machine with Intel Core i7-7700 Processor (3.6GHz), and took 0.0089, 0.0077, and 0.0083 seconds per iteration, respectively.

Figures~\ref{fig:sna_pmf} to \ref{fig:sna_point_est} show some key inference results, except for the trace plots of the parameters, which are in the online Appendix D.2. Of more importance is the posterior density (or mass function in the case of $K_n$ and $k$) in Figures~\ref{fig:sna_pmf} and \ref{fig:sna_pdf}. On one hand, in the panels for $k$ and $\gamma$, the infinite regime is naturally missing, while the results for the finite regime and model selection coincide as expected. Similarly, in the panels for $\theta$ and $\alpha$, the finite regime is naturally missing, while the results for the infinite regime and model selection coincide as expected. On the other hand, the panels for $K_n$, $a$ and $b$ illustrate that the posterior density from model selection is essentially a weighted average of that of the two regimes.

The departure of the posterior densities of $\alpha$ and $\gamma$ from 0 suggests that either the infinite regime or the finite regime is preferred to their shared boundary, i.e. the Dirichlet process when $\alpha=\gamma=0$. Between the two regimes, with $\P(r=1)=0.2$ resulting in $\P(r=1|\Y)=0.5474$, the Bayes factor $B_{10}=\displaystyle\frac{\P(r=1|\Y)}{\P(r=0|\Y)}\left/\frac{\P(r=1)}{\P(r=0)}\right.=4.8368$, suggesting a slight preference for the finite regime, for this citation network.

\begin{knitrout}
\definecolor{shadecolor}{rgb}{0.969, 0.969, 0.969}\color{fgcolor}\begin{figure}[!ht]

{\centering \includegraphics[width=0.85\linewidth]{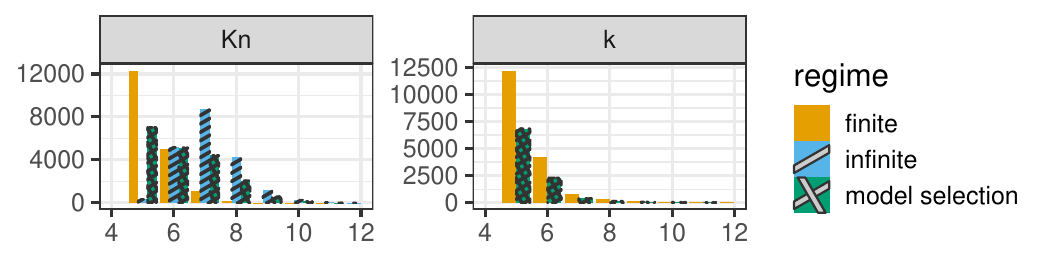} 

}

\caption[Posterior histogram of $K_n$ and $k$, for the finite (solid) and infinite (stripe) regimes of the DAG-SBM, and model selection (crosshatch), for the SNA citation network]{Posterior histogram of $K_n$ and $k$, for the finite (solid) and infinite (stripe) regimes of the DAG-SBM, and model selection (crosshatch), for the SNA citation network. The right panel is for finite regime and model selection only as $k$ does not exist in the infinite regime. For readability, values of $k$ above $12$ are not shown.}\label{fig:sna_pmf}
\end{figure}

\end{knitrout}
\begin{knitrout}
\definecolor{shadecolor}{rgb}{0.969, 0.969, 0.969}\color{fgcolor}\begin{figure}[!ht]

{\centering \includegraphics[width=0.85\linewidth]{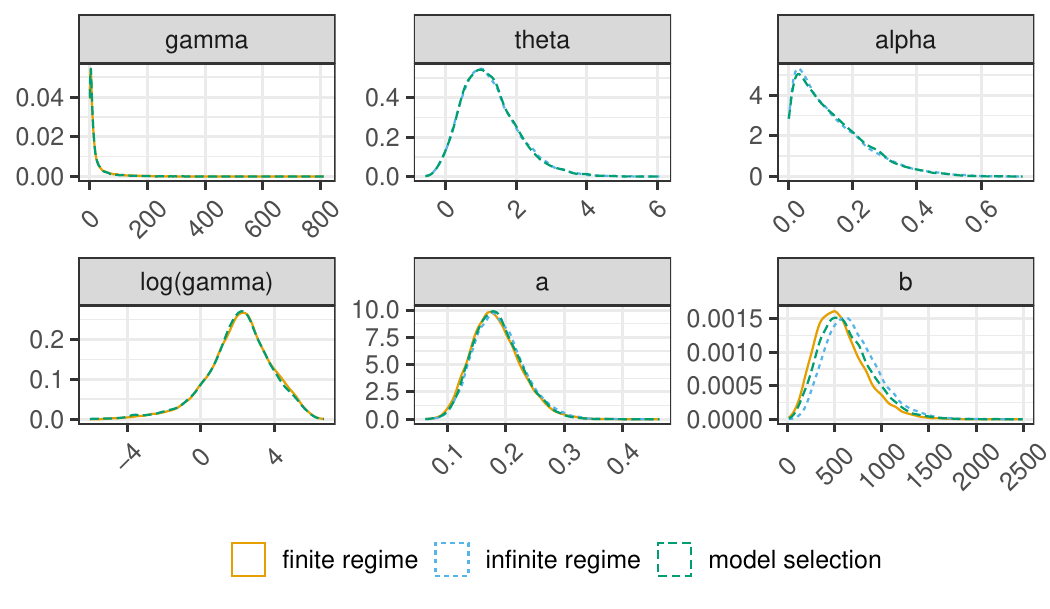} 

}

\caption[Posterior density of the parameters for the infinite (dotted) and finite (solid) regime of the DAG-SBM, and model selection (dashed), for the SNA citation network]{Posterior density of the parameters for the infinite (dotted) and finite (solid) regime of the DAG-SBM, and model selection (dashed), for the SNA citation network.}\label{fig:sna_pdf}
\end{figure}

\end{knitrout}
\begin{knitrout}
\definecolor{shadecolor}{rgb}{0.969, 0.969, 0.969}\color{fgcolor}\begin{figure}[!htpb]

{\centering \includegraphics[width=0.52\linewidth]{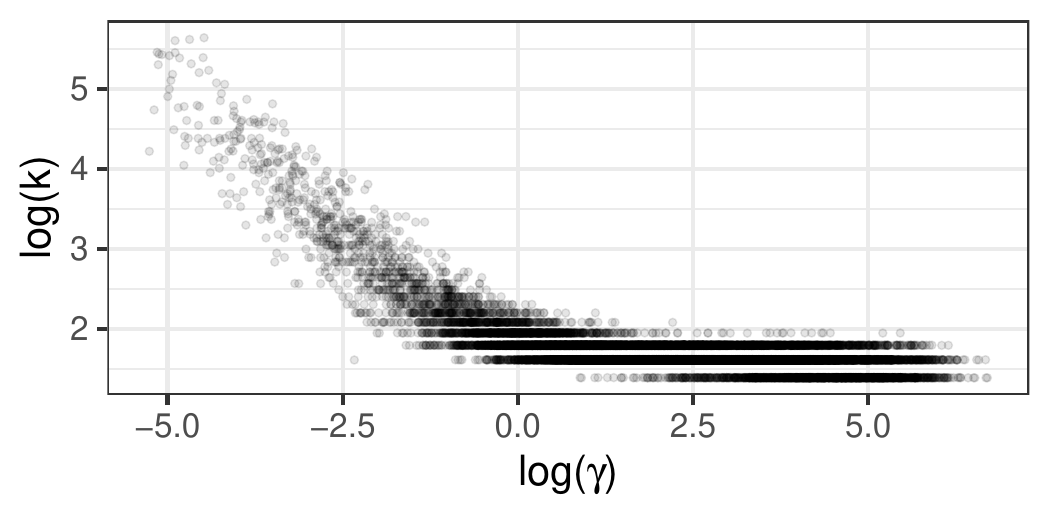} 

}

\caption[Joint posterior of $\log \gamma$ and $\log k$ for the finite regime of the DAG-SBM, for the SNA citation network]{Joint posterior of $\log \gamma$ and $\log k$ for the finite regime of the DAG-SBM, for the SNA citation network.}\label{fig:sna_k_gamma}
\end{figure}

\end{knitrout}

\begin{knitrout}
\definecolor{shadecolor}{rgb}{0.969, 0.969, 0.969}\color{fgcolor}\begin{figure}[!hpt]

{\centering \includegraphics[width=0.45\linewidth]{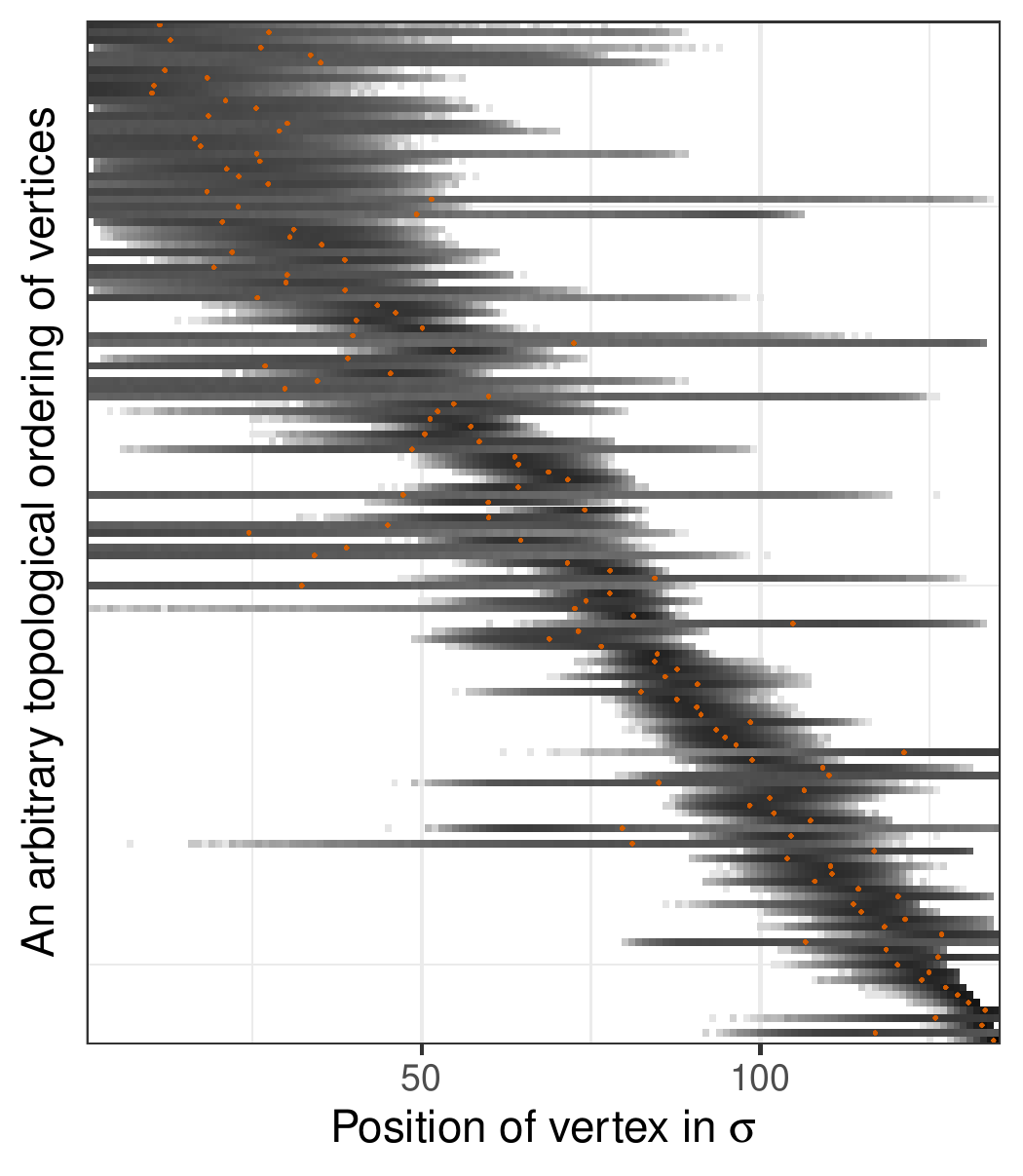} 
\includegraphics[width=0.45\linewidth]{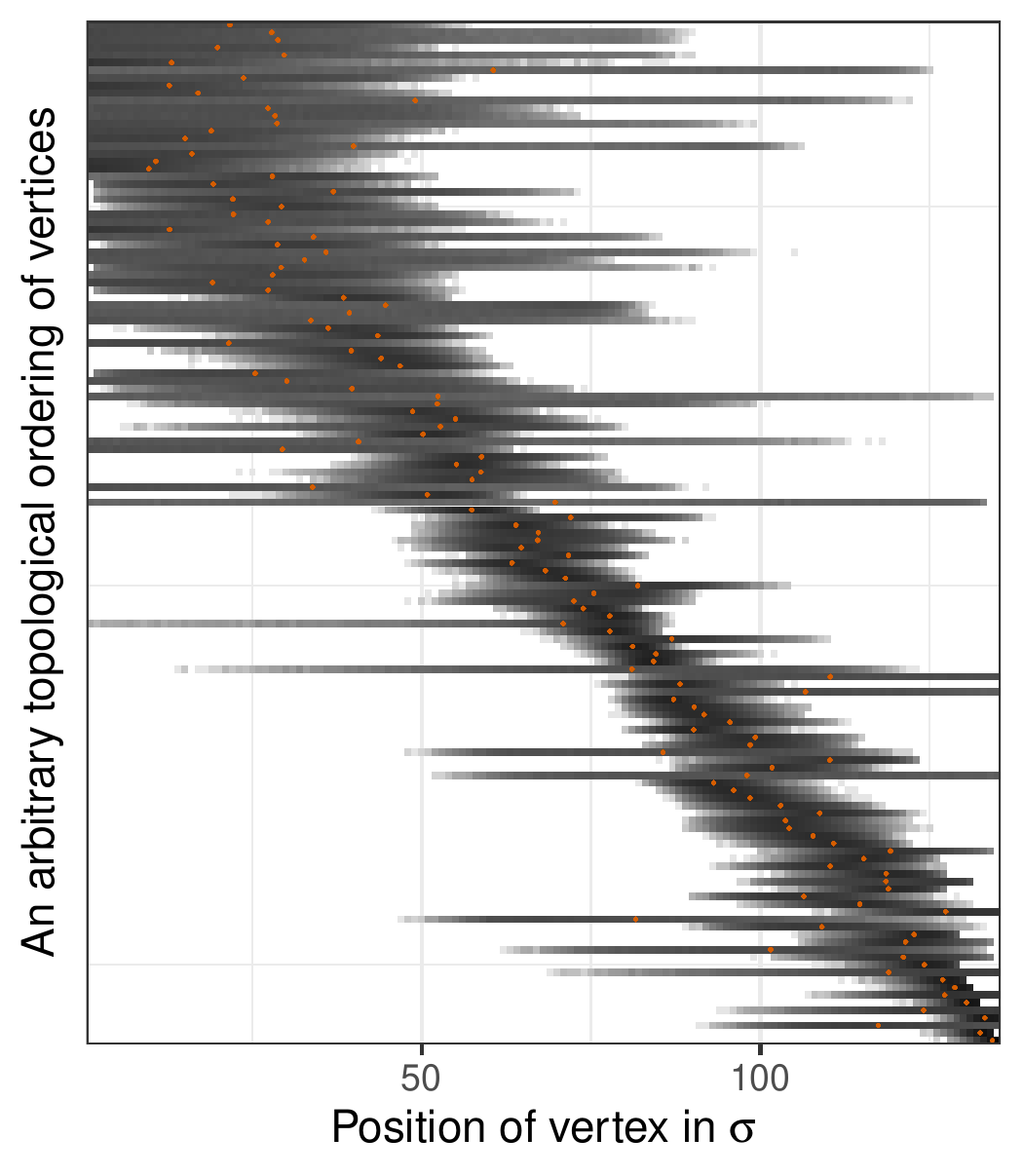} 

}

\caption[The posterior density of the positions of the vertices in $\bsigma$ for the infinite (left) and finite (right) regime of the DAG-SBM, for the SNA citation network]{The posterior density of the positions of the vertices in $\bsigma$ for the infinite (left) and finite (right) regime of the DAG-SBM, for the SNA citation network. Each row represents the posterior density of a component of $\p$. The coloured dots are the mean positions.}\label{fig:sna_topo}
\end{figure}

\end{knitrout}
\begin{knitrout}
\definecolor{shadecolor}{rgb}{0.969, 0.969, 0.969}\color{fgcolor}\begin{figure}[!ht]

{\centering \includegraphics[width=0.45\linewidth]{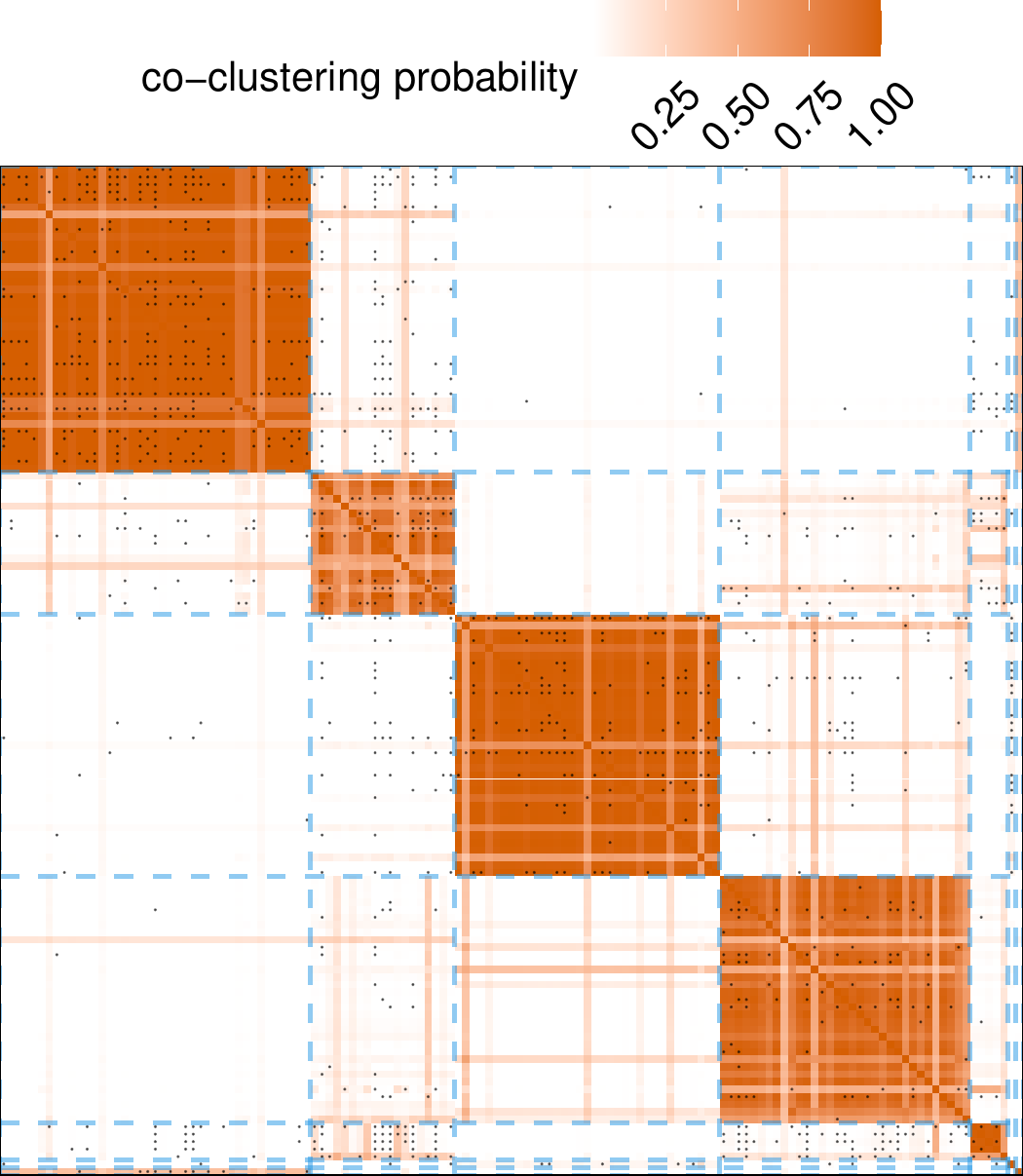} 
\includegraphics[width=0.45\linewidth]{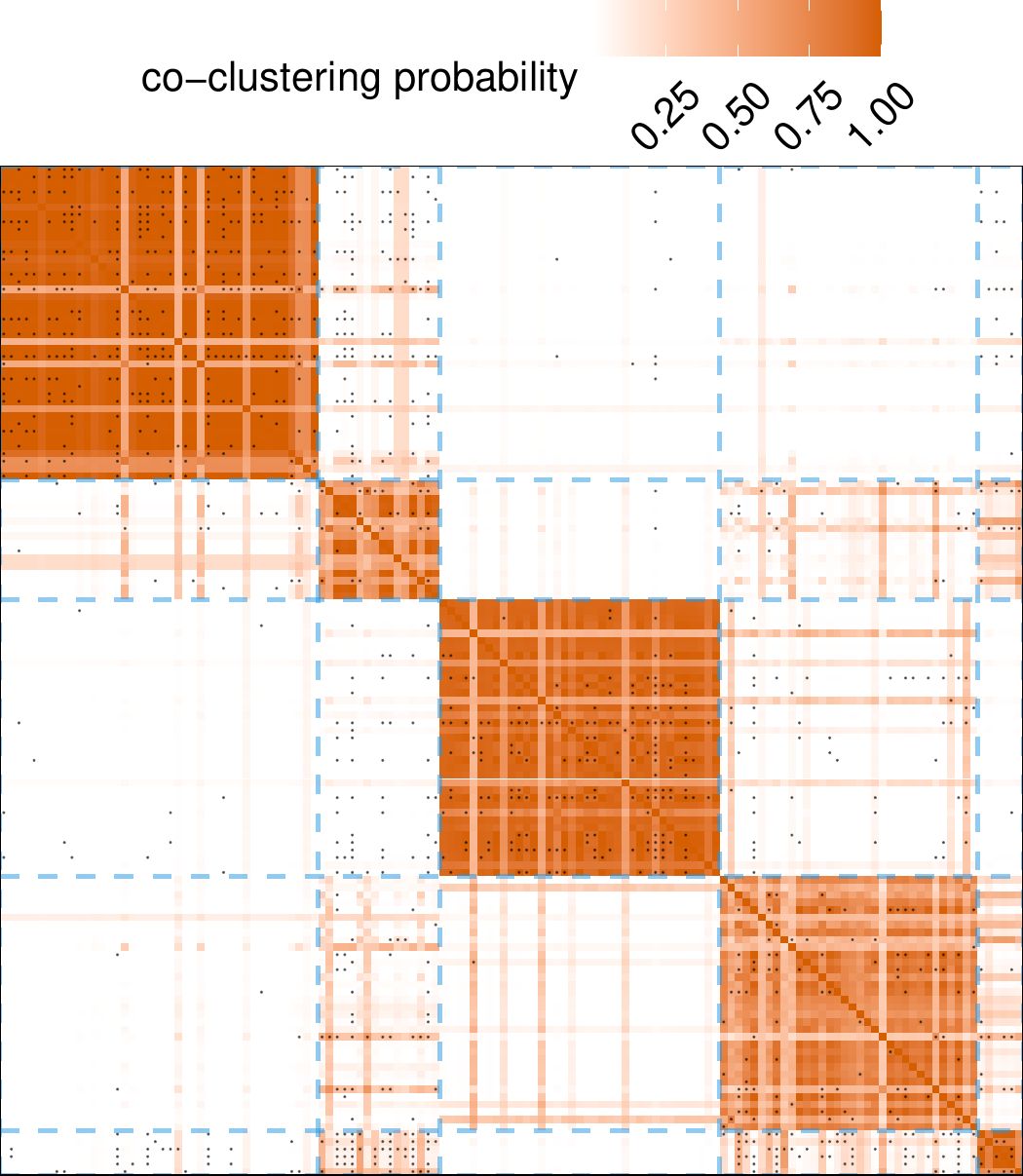} 

}

\caption[The co-clustering matrix (spectrum) and the adjacency matrix (dots) for the infinite (left) and finite (right) regime of the DAG-SBM, for the SNA citation network]{The co-clustering matrix (spectrum) and the adjacency matrix (dots) for the infinite (left) and finite (right) regime of the DAG-SBM, for the SNA citation network. The vertices are clustered (dashed lines) according to the point estimate $\hat{\Z}_n$.}\label{fig:sna_point_est}
\end{figure}

\end{knitrout}

The high number of thinning is mainly due to the Markov chain for the finite regime occasionally getting stuck at $K_n=4$, which is in turn due to the mixing of $k$ and $\gamma$, and the skewness of their joint posterior, which is plotted on log scale in Figure~\ref{fig:sna_k_gamma}. There is a dense cloud of points towards the bottom right of the plot, indicating high posterior density around small values of $k$ (and high values of $\gamma$), thus limiting $K_n$ to grow. However, such issue of getting stuck at small values of $k$ and $K_n$ does not exist for either the infinite regime or the application to the larger dataset in the next subsection. 

Next, we look at the posterior of $\bsigma$, or equivalently the positions of the vertices in the topological ordering, $\p$. The mixing of the MCMC is good, and the trace plots for some components of $\p$ are shown in the online Appendix D.2. The posterior density of each component of $\p$ is plotted as a row in Figure~\ref{fig:sna_topo}, with the rows themselves in an arbitrary topological ordering. As the citations, i.e. the edges in the DAG in general go from more recent works to older ones, the top (bottom) rows, which correspond to topologically earlier (later) vertices, are generally the more recent (older) articles in the network. The coloured part of each row represents the support of the posterior of the corresponding article, and in general is narrower towards the bottom. An interpretation is that generally chronologically older works are likely to cite between themselves and get cited by later ones, thus limiting their positions in $\bsigma$. Lastly, using Equation~\ref{eq:point.estimate} and the SALSO algorithm \citep{Dah(21)} to minimise the loss function, the point estimate $\hat{\Z}_n$ is obtained, with the co-clustering and adjacency matrices plotted in Figure~\ref{fig:sna_point_est}. %and the network diagram in Figure \ref{fig:sna_network}.

\subsection{Statistics citation network} \label{sec:citation}

\begin{knitrout}
\definecolor{shadecolor}{rgb}{0.969, 0.969, 0.969}\color{fgcolor}\begin{figure}[!hb]

{\centering \includegraphics[width=0.58\linewidth]{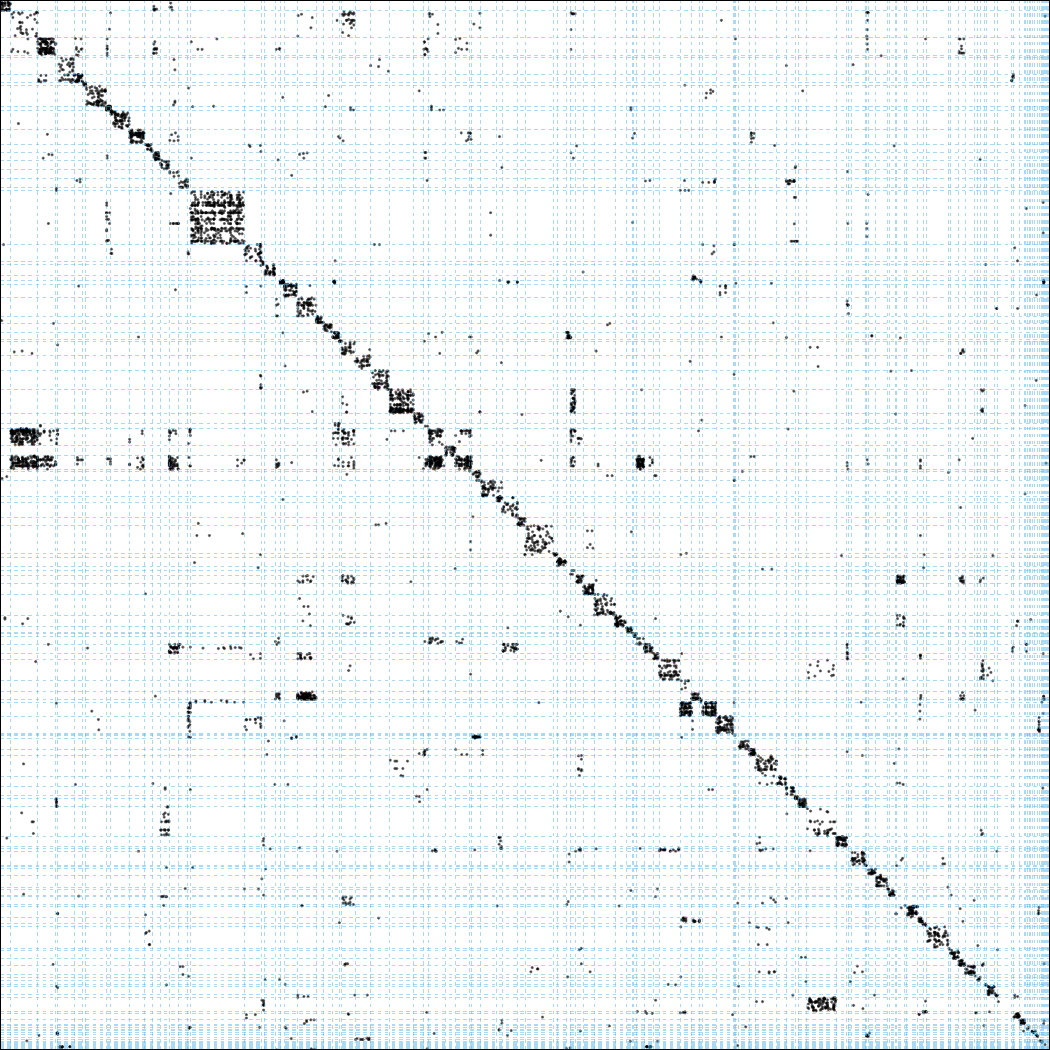} 

}

\caption[The adjacency matrix (black dots) for the finite regime, for the statistics citation network]{The adjacency matrix (black dots) for the finite regime, for the statistics citation network. The vertices are clustered (dashed lines) according to the point estimate $\hat{\Z}_n$.}\label{fig:jaj_point_est}
\end{figure}

\end{knitrout}

We apply the model to a second citation network. The original data analysed by \citet{jj16} contains 5722 citations between 3248 articles in the top statistics journals, from 2003 to the first half of 2012. We only consider the largest connected component, and remove one of the two edges in every pair of cyclic citations (there were only 9 pairs of these). Upon the data cleaning, we arrive at a citation network that is a DAG and contains 5563 citations between 2248 articles. We refer to this data as the statistics citation network hereafter.

The MCMC sampler was applied with \ensuremath{1.5\times 10^{4}} iterations obtained after a burn-in period of \ensuremath{5.5\times 10^{4}} and thinning of 10, with the same priors for $a$, $b$, $\alpha$, $\theta$, $k$ and $\gamma$ as those in Section~\ref{sec:real.data}. For model selection, even with $\P(r=0)=0.99$, the whole chain stays in the finite regime, i.e. $\P(r=1|\Y)=1$. Therefore, we shall report the results under the finite regime only, as it is heavily preferred to the infinite counterpart for this network. The trace plots and posterior densities are provided in the online Appendix D.3. The posterior of $K_n$ ranges from 92 to 124, which seems reasonable with around 30 articles per cluster.

Similar to Figure~\ref{fig:sna_point_est}, the adjacency matrix for the statistics citation network is plotted in Figure~\ref{fig:jaj_point_est}, with the vertices clustered according to the point estimate. The co-clustering matrix is not being overlaid here to preserve the figure readability. The concentration of the black dots along the major block diagonal suggests that most groups are closely knitted. On the other hand, there are some concentrated blocks which are off-diagonal and asymmetric, indicating high number of one-way citations from one group to another. 

\begin{table}[!b]
\small
\caption{\label{tab:appendix_jaj_group_A}Titles of top articles  according to topological ordering for selected groups.}
\centering
\begin{tabular}[t]{l|l}
\hline
Group & Article\\
\hline
18 ``False Discovery'' & Genovese and Wasserman (AoS, 2004)\\
\hline
18 ``False Discovery'' & Storey (AoS, 2003)\\
\hline
18 ``False Discovery'' &  Johnstone (AoS, 2008)\\
\hline
18 ``False Discovery'' &  Black (JRSSB, 2004)\\
\hline
18 ``False Discovery'' & Cox and Wong (JRSSB, 2004)\\
\hline
45 ``Design and Randomization'' & Lockwood, Schervish, Gurian and  Small (JASA, 2004)\\
\hline
45 ``Design and Randomization'' & Montanari and Ranalli (JASA, 2005) \\
\hline
45 ``Design and Randomization'' &  Rubin (JASA, 2005)\\
\hline
45 ``Design and Randomization'' &  Barnard, Frangakis, Hill and Rubin (JASA, 2003) \\
\hline
45 ``Design and Randomization'' &  Greevy,  Silber,  Cnaan and  Rosenbaum  (JASA, 2004) \\
\hline
2 ``Wavelet techniques'' &  Averkamp and Houdré (AoS, 2005)\\
\hline
2 ``Wavelet techniques'' & Morris,  Vannucci,  Brown and Carroll (JASA, 2003)  \\
\hline
2 ``Wavelet techniques'' &  Yu and Jones (JASA, 2004)\\
\hline
2 ``Wavelet techniques'' & Foster and Stine (JASA, 2004)\\
\hline
2 ``Wavelet techniques'' & Signorini and Jone (JASA, 2004) \\
\hline
71 ``Missing Data'' &  Rao, Mingo, Speicher and Edelman (AoS, 2008)\\
\hline
71 ``Missing Data'' &  Li,  Aragon, Shedden and Agnan (JASA, 2003)\\
\hline
71 ``Missing Data'' &    Stute, Xue and Zhu (JASA, 2007) \\
\hline
71 ``Missing Data'' &  Liang, Wang, Robins and  Carroll (JASA, 2004) \\
\hline
71 ``Missing Data'' &  Chen, Leung and Qin (JASA, 2003) \\
\hline
25 ``Functional Regression'' &  Crambes, Kneip and Sarda (AoS, 2009) \\
\hline
25 ``Functional Regression'' &  Hall and Horowitz (AoS, 2007)\\
\hline
25 ``Functional Regression'' &   Cai and Hall (AoS, 2006)\\
\hline
25 ``Functional Regression'' &  Wu and Liu (JASA, 2007)\\
\hline
25 ``Functional Regression'' &  Mammen and Nielsen (Biometrika, 2003) \\
\hline
65 ``Spatial Processes'' &  Wang and Carey (JASA, 2004) \\
\hline
65 ``Spatial Processes'' &    Schmidt and O'Hagan (JRSSB, 2003) \\
\hline
65 ``Spatial Processes'' &   Schlather, Ribeiro Jr and Diggle (JRSSB, 2004)\\
\hline
65 ``Spatial Processes'' &  McElroy and Politis (AoS, 2007)\\
\hline
65 ``Spatial Processes'' &  Gelfand, Kim, Sirmans and Banerjee (JASA, 2003) \\
\hline
\end{tabular}
\end{table}

We report several groups of top articles according to median topological ordering for some large groups in Table~\ref{tab:appendix_jaj_group_A}. Specifically, for a selection of blocks, we report the top 5 articles in topological ordering within the block. From the article titles, groups seem to roughly correspond to various topics in statistics. For example, group 18 seems to cover false discovery rates, group 45 design and randomisation, group 2 wavelets, group 71 missing/perturbed data, group 25 functional linear regression, and group 65 spatial processes. However, note that the reported group memberships are just point estimates obtained from the MCMC draws, and therefore there might be some possibly large posterior uncertainty in the topological ordering for some of these articles.

Lastly, we report in Table~\ref{tab:appendix_jaj_group_B} some top articles according to the median topological ordering, this time without the group memberships. In general, these articles have high number of citations. However, we also notice that some of these articles might not have a very high number of citations. This is due to the fact that these articles are cited by some other influential articles towards the end of the topological ordering.

\begin{table}[!b]
  \small
\caption{\label{tab:appendix_jaj_group_B}Titles of top 20 articles according to topological ordering.}
\centering
\begin{tabular}[t]{|l|}
\hline
Article \\
\hline
Johnstone (AoS, 2008) \\
\hline
Genovese and Wasserman  (AoS, 2005) \\
\hline
Zhang, Siegmund, Ji and Li (Biometrika, 2010) \\
\hline
Averkamp and Houdré (AoS, 2005) \\
\hline
Ishwaran and  Rao (JASA, 2003) \\
\hline
Zeng (AoS, 2004) \\
\hline
Beskos, Papaspiliopoulos and Roberts (AoS, 2009) \\
\hline
 Schlather and Tawn (Biometrika, 2003) \\
\hline
Mammen and Nielsen (Biometrika, 2003) \\
\hline
Wang and Carey (Biometrika, 2003) \\
\hline
Zhang (JASA, 2003) \\
\hline
Huang (AoS, 2004) \\
\hline
Hu and Rosenberger (JASA, 2003) \\
\hline
Ramsay, Hooker, Campbell and Cao (JRSSB, 2007) \\
\hline
Storey (AoS, 2003) \\
\hline
Lee and Pun (JASA, 2006) \\
\hline
Zhu, Miao and Peng (JASA, 2006) \\
\hline
Barnard, Frangakis, Hill and Rubin (JASA, 2003) \\
\hline
Foster and Stine (JASA, 2004)\\
\hline
Lue (Biometrika, 2004)\\
\hline
\end{tabular}
\end{table}

\section{Discussion} \label{sect.discussion}
In this article, we proposed a Bayesian nonparametric SBM for DAGs. By conditioning on a latent topological ordering, the likelihood of the data (which is composed of directed edges) becomes that of an upper diagonal adjacency matrix. The topological ordering is treated as an unknown parameter, endowed with a prior and inferred a posteriori within the MCMC sampler, using a modified leap-and-shift proposal, previously used for the ranking in the Mallows model by \cite{vscfa18}. The use of the PY process prior for the allocation vector $\Z[n]$ allows the model to infer the number of groups $K_n$ from the data. Moreover, a model selection step for the two regimes of the PY process (one in which the total number of blocks is finite and unknown, and another one in which it increases with the sample size) is incorporated in the MCMC sampler. 
%The model and sampler are applied to two citation networks.

The model can be generalised in different ways. For example, it can be extended by introducing covariate information, such as the keywords of each document or its publication year. This could be achieved by modelling the degree correction factors with a covariate dependent distribution. Also, in terms of parametrisation, the two regimes of the PY process could be unified so that $\gamma$ and $\alpha$ become one parameter that can take a value between $-\infty$ and $1$, with its posterior density directly implying which regime is preferred. The main obstacle to overcome here would be the sampling from the non-standard joint parameter space of $\theta$ and $\alpha$ across the two regimes. Another issue to be resolved is the inference of $k$, which is naturally highly correlated with $K_n$, under the finite regime. This is apparent in the parameter trace plots (in online Appendix D.3) for the statistics citation network, while for the SNA citation network the model selection improves the mixing of $k$ in the MCMC. Ideally $k$ is integrated out, but the computations required mean that this is feasible only under certain special cases. Such issue with $k$ remains to be resolved.

There are potential extensions regarding inference procedure and results. Similar to how $\hat{\Z}_n$ is computed for $\Z[n]$, a point estimate could be provided for $\bsigma$, but the distance function for the ordering has to be carefully considered. Relatedly, a Mallow's model prior \citep{vscfa18} could be used for $\bsigma$, as opposed to the uniform prior used here, to potentially provide more information to facilitate the inference. Lastly, the derivation of an efficient Variational Bayes algorithm \citep{ble17} would possibly allow the applicability of the proposed model to much larger datasets.

\section*{Funding Acknowledgements}
This research received no specific grant from any funding
agency in the public, commercial, or not-for-profit sectors.

\section*{Disclosure Statement}
The authors report there are no competing interests to declare.

\section*{Supplementary Materials}

A single archive, available online, contains all of the following supplementary materials:

\begin{description}

\item[Appendices:] Background materials of SBM and PY process, likelihood derivations, MCMC algorithms, and additional plots (\textbf{appendices.pdf}, pdf file).
\item[R package for simulations \& MCMC sampler:] R-package ``dagsbm'' containing code to run the MCMC sampler for the DAG-SBM, and to simulate graphs from the model. The package also contains the dataset of SNA citation network presented in the article. (\textbf{dagsbm\_0.0.8.tar.gz}, GNU zipped tar file)
\item[Code:] Script for running the MCMC sampler using the R package. (\textbf{sim\_dc.R}, R script)
\item[Results:] Pre-generated results from running sim\_dc.R, for the README and reproducibility checks. (\textbf{results/*sim\_dc\_alpha=0.00\_theta=1.00\_n=250*.rds}, Rds file)
\item[README:] Instructions for reproducing the results in the paper using the R package and code. (\textbf{README.Rmd}, R Markdown script)
  
\end{description}

\bibliographystyle{agsm}
\bibliography{ref_nr}

\end{document}